\documentclass[universe,review,submit,pdftex,moreauthors,dvipsnames]{Definitions/mdpi} 
\preto{\abstractkeywords}{\nolinenumbers} % remove line number

\firstpage{1} 
\pubvolume{1}
\issuenum{1}
\articlenumber{0}
\pubyear{2023}
\copyrightyear{2023}
\datereceived{ } 
\daterevised{ } % Comment out if no revised date
\dateaccepted{ } 
\datepublished{ } 
\hreflink{https://doi.org/} % If needed use \linebreak
\usepackage[para,online,flushleft]{threeparttable}
\usepackage{txfonts}
\usepackage{siunitx}
\usepackage[normalem]{ulem}
\usepackage{subcaption}

\Title{Measuring the Lense–Thirring Orbital Precession and the
Neutron Star Moment of Inertia with Pulsars}
\TitleCitation{Measuring the Lense–Thirring Orbital Precession and the
Neutron Star Moment of Inertia with Pulsars}

\Author{Huanchen Hu~* \orcidA{} and Paulo C. C. Freire *\orcidB{}}

\AuthorNames{Huanchen Hu and Paulo C. C. Freire}

\AuthorCitation{Hu, H. and P. C. C. Freire}

\address[1]{%
Max-Planck-Institut f\"ur Radioastronomie, Auf dem H\"ugel 69, D-53121 Bonn, Germany}
\corres{\hangafter=1 \hangindent=1.05em \hspace{-0.82em}Correspondence: huhu@mpifr-bonn.mpg.de (H.H.); pfreire@mpifr-bonn.mpg.de (P.C.C.F.)}

\abstract{
Neutron stars (NSs) are compact objects that host the densest forms of matter in the observable universe, providing unique opportunities to study the behaviour of matter at extreme densities.
While precision measurements of NS masses through pulsar timing have imposed effective constraints on the equation of state (EoS) of dense matter, accurately determining the radius or moment of inertia (MoI) of a NS remains a major challenge.
This article presents a detailed review on measuring the Lense-Thirring (LT) precession effect in the orbit of binary pulsars, which would give access to the MoI of NSs and offer further constraints on the EoS. 
We discuss the suitability of certain classes of binary pulsars for measuring the LT precession from the perspective of binary star evolution, and highlight five pulsars that exhibit properties promising to realise these goals in the near future.
Finally, discoveries of compact binaries with shorter orbital periods hold the potential to greatly enhance measurements of  the MoI of NSs. The MoI measurements of binary pulsars are pivotal to advancing our understanding of matter at supranuclear densities as well as improving the precision of gravity tests, such as the orbital decay due to gravitational wave emission and of tests of alternative gravity theories.
}


\begin{document}

%%%%%%%%%%%%%%%%%%%%%%%%%%%%%%%%%%%%%%%%%%

%\tableofcontents

%%%%%%%%%%%%%%%%%%%%%%%%%
\section{Introduction}

\subsection{The equation of state of dense matter}\label{sec:intro1}

\begin{figure}[t]
    \centering
    \includegraphics[width=0.95\textwidth]{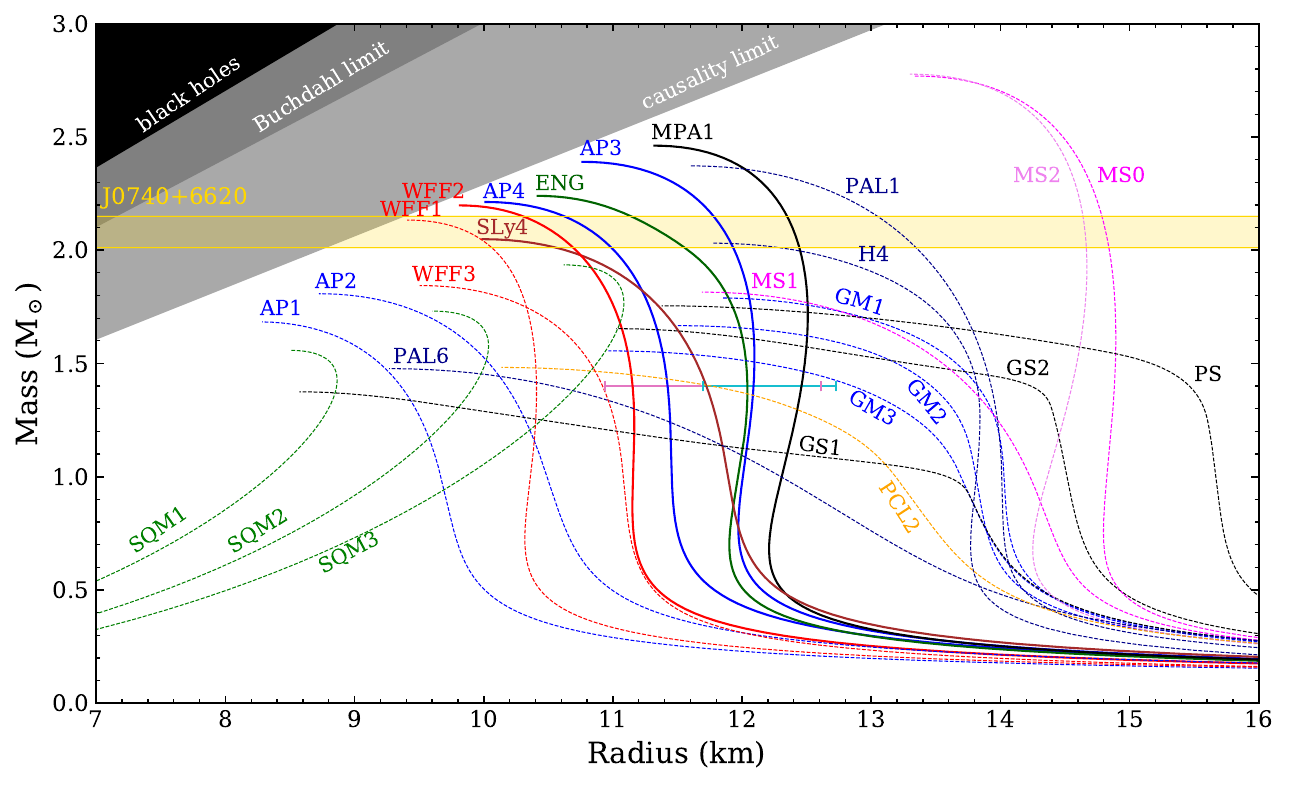}
    \caption{The NS mass-radius relation for different EoSs listed in \cite{Lattimer_2001}. The horizontal lines in yellow represents the 1-$\sigma$ mass ranges for the most massive NS known, PSR~J0740+6620 \cite{Fonseca+2021}. The pink bar in the middle shows the multimessenger constraint on the radius of a 1.4-$\mathrm{M_\odot}$ NS \cite{Dietrich+2020Sci} at 90\% confidence, and the cyan bar shows an updated radius constraint at 95\% confidence \cite{Koehn+2024}. The EoSs plotted in dashed lines are excluded by the lower limit of the maximum NS mass and the constraint of radius range in \cite{Dietrich+2020Sci}. Figure courtesy of Norbert Wex.}
    \label{fig:RM}
\end{figure}

Since the time of the first detailed calculation of the structure of neutron stars (NSs), in 1939 \citep{OppenheimerVolkoff1939}, it has been recognised that the main uncertainty regarding their bulk properties (mass and maximum mass possible, radius, tidal deformability, moment of inertia (MoI) and quadrupolar moment) is lack of knowledge of the microscopic properties of cold nuclear matter at the densities one finds in NS cores, which are well above that of the atomic nucleus. Especially important for the calculation of the NS structure is the relation between pressure and density of this type of matter --- its equation of state (EoS). 
The EoS had large uncertainties in 1939 and, to some extent, these uncertainties persist today \cite{Lattimer_2001, LIGO2018,Lattimer2019Univ,Capano2020,Dietrich+2020Sci}. This results in a wide range of NS mass-radius relations (see Fig.~\ref{fig:RM}).

This situation happens for several reasons, all related to the properties of the strong nuclear force: a) Given the gluon self-interaction, perturbative methods do not allow precise theoretical calculations of the behaviour of the strong nuclear force, in particular the state and composition of matter at high densities; and b) Laboratory experiments, in particular ion-ion collisions, can reach similar densities to those found in NSs, but for very short amounts of time and with very high kinetic energies per nucleon, a situation that is very different from what one finds in NS cores \cite{Lattimer_2001}. 

For this reason, the measurement (via astronomical observations) of bulk NS properties plays an important role in constraining the EoS and understanding the state of cold, dense neutron matter, which is a major research problem in fundamental physics and astrophysics \cite{OzelFreire2016}. 

The most straightforward way to constrain the dense matter EoS is via the measurement of NS masses. These masses can be measured through relativistic orbital effects in binary pulsars using pulsar timing, they can also be potentially combined with optical information from the companion in case of a white dwarf (WD). 
In particular, the central matter must be incompressible enough (i.e., the EoS must be "stiff" enough) to support the maximum observed NS mass.
The most massive pulsar, PSR~J0740+6620  ($m_\mathrm{p}=2.08\pm0.07\,\mathrm{M_\odot}$,\footnote{See Section~\ref{sec:SOG} for the definition of the nominal solar mass $\mathrm{M_\odot}$.} \cite{Fonseca+2021}) has excluded a number of soft EoSs that can not support a 2-$\mathrm{M_{\odot}}$ NS (see Fig.~\ref{fig:RM}). There are hints of more massive NSs from measurements of stellar companions other than WDs (PSR~J0952$-$0607 \cite{Romani+2022ApJ} and PSR~J2215+5135 \cite{Linares+2018ApJ}), but these mass measurements depend, to a large extent, on the interpretation of stellar spectra and on detailed models of stellar surfaces, and have given divergent results in the past (e.g., \cite{vanKerkwijk+2011,Clark+2023NatAs}).
Therefore, as we can see in Fig.~\ref{fig:RM}, quite a few EoSs survive. The reason is that this method only probes one dimension of the parameter space shown in Fig.~\ref{fig:RM}, as we see, the surviving EoSs have a wide range of radii. This means that any method where two bulk parameters can be probed simultaneously for the same NS, such as mass and radius or mass and MoI, would represent an ever more powerful constraint.

Measuring the extremely small radii of NSs ($\sim 10$~km) is a challenging task.
Given the small size of NSs (radii $\sim 10$~km), it is quite challenging to accurately measure their radii. This has been attempted using a wide variety of methods with X-ray observations \cite{OzelFreire2016,Ascenzi+2024}, such as pulse profile modelling using Neutron Star Interior Composition Explorer (NICER) \cite{NICER,Riley2019, Miller2019,Riley+2021ApJ,Vinciguerra+2024}.
These methods provide valuable radius measurements, and are even more precise in cases where the mass is measured independently from radio timing measurements, e.g. PSR J0740+6620, a rare high-mass NS that will likely not be so often present in GW mergers. However, they depend on very detailed and difficult modelling of X-ray light curves.

A more recent method uses LIGO's detailed gravitational wave (GW) measurements of the GW170817 NS-NS merger \cite{LIGO2017}. The resulting estimates of tidal deformability have excluded the very stiffest EoSs considered by \cite{LIGO2018}, including the MS1, MS1b and H4 EoSs, as they would cause observable tidal effects in the GW waveform that are not consistent with the observation. This method has great potential for the future, with more sensitive GW detectors and louder events, but for the moment the constraints have rather large uncertainties. 

\begin{figure}[t]
    \centering
    \includegraphics[width=0.8\textwidth]{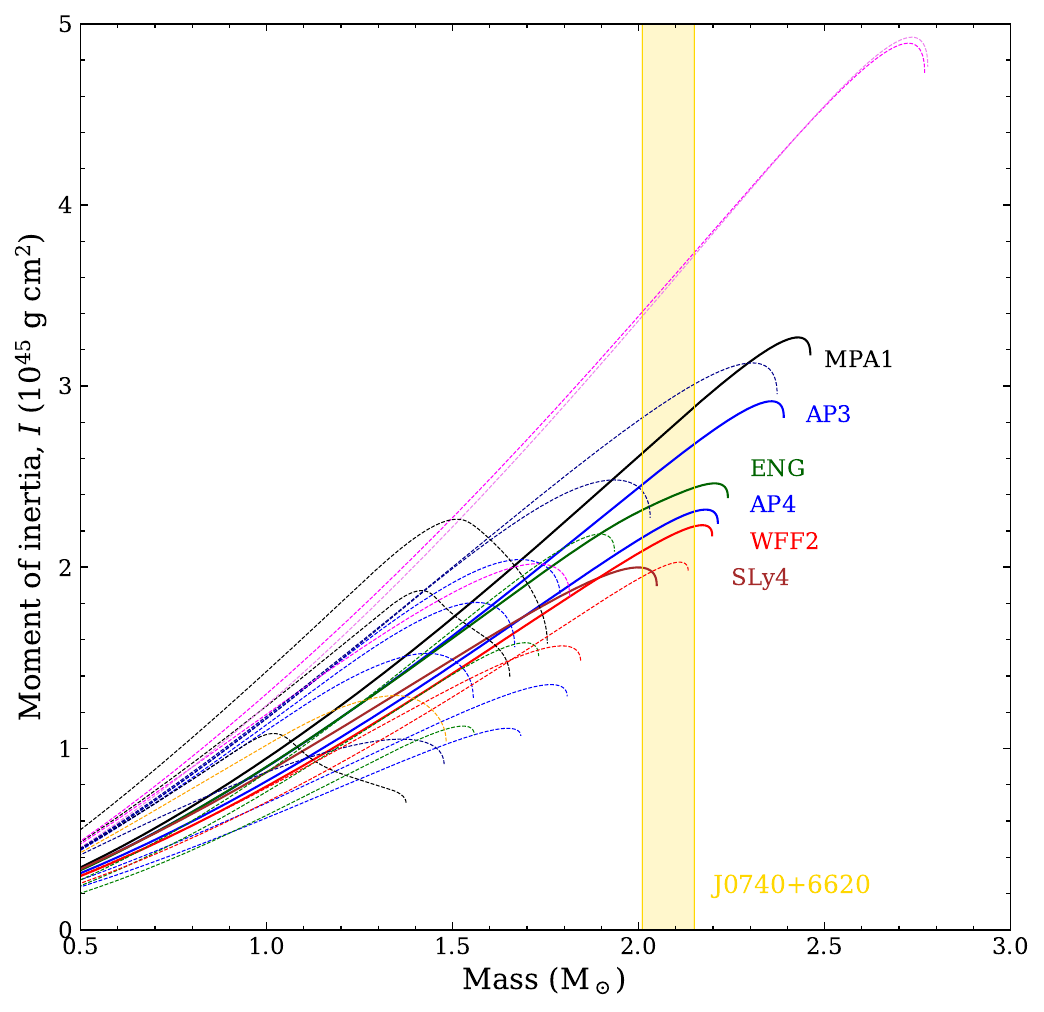}
    \caption{The change of the MoI with mass for different EoSs, coloured as in Fig~\ref{fig:RM}. The yellow band represents the mass range of PSR~J0740+6620. The EoSs marked with labels survive the current lower limit of maximum NS mass and multimessenger radius constraint.
    Figure courtesy of Norbert Wex.}
    \label{fig:M_MOI}
\end{figure}

The method we will discuss here is to measure the MoI of binary pulsars via very small effects of relativistic spin-orbit coupling on their orbital motion, the {\em Lense-Thirring (LT) precession}, which was first suggested by Damour \& Sch\"afer in 1988 \cite{DS88}. 
Using radio pulsar timing (and our knowledge of the system geometry), we attempt to measure simultaneously the mass {\em and} the MoI of a NS. This is in principle equivalent to measuring simultaneously mass and radius, because, as we can see in Fig.~\ref{fig:M_MOI}, each EoS gives a unique prediction for the relation between mass and MoI. However, it might in effect be even more valuable, as it would provide direct constraints on the tidal deformability measured by ground-based GW detectors and the quadrupole moment of NSs via the universal (EoS-independent) I-Love-Q relation \cite{I-Love-Q}.

Measuring the MoI can be done because of the high precision of pulsar timing. This method is intrinsically {\em accurate}, because the timing techniques are model-independent. 
In addition, as the MoI of NS relates to its mass and radius as $I_\mathrm{NS}^{} \propto M_\mathrm{NS}^{} R_\mathrm{NS}^2$, the range of MoI for different EoS models is relatively larger than the range of radius (see Fig.~\ref{fig:M_MOI}, \cite{LS05}).
Also, the relative accuracy required for the MoI measurement is less than that of radius as $\Delta I_\mathrm{NS}^{} / I_\mathrm{NS}^{} = 2\, \Delta R_\mathrm{NS}/R_\mathrm{NS}$. 
These make the measurement of the MoI easier than that of radius.
However, they are not yet {\em precise}, because of the small magnitude of the effects being measured.

This article aims to provide a review on measuring the MoI of a NS using LT precession with the latest knowledge on promising pulsar systems. 
In Section~\ref{sec2}, we provide a brief overview of the LT precession and how it can be tested using pulsar timing. 
In Section~\ref{sec:evo}, we explain the evolution of different types of binary pulsar systems and the suitability of double neutron star (DNS) systems for testing the LT precession.
Several binary pulsars are discussed in detail in Section~\ref{sec:psrs}, including PSR~J1141$-$6545, PSR~J1757$-$1854, the Double Pulsar (PSR~J0737$-$3039A) and PSR~J1946+2052. 
A discussion on the recent study of the massive binary pulsar J0514$-$4002E is given in Section~\ref{sec:J0514}.
We discuss the future discoveries of ultra compact DNS systems and the significance of LT precession measurements in relation to orbital period and eccentricity in Section~\ref{sec:discovery}.
Finally, in Section~\ref{sec:sum}, we provide a summary and outlook for the future.

However, before we move on, we must say a few words about geodetic precession, which has been observed in binary pulsars and is intimately linked with LT precession.

%%%%%%%%%%%%%%%%%%%%%%%%%%%%%%
\subsection{Spin-orbit coupling and geodetic precession}
\label{sec:SOG}

The coupling of the orbital and spin dynamics produces two observable effects. The change of the direction of the spin is known as {\em geodetic precession}. The corresponding change of direction of the orbital angular momentum, which is necessary to conserve the total angular momentum of the system, is the LT precession. The magnitudes of these vectors stay the same \cite{Barker1975ApJ}.

\begin{figure}[t]
    \centering
    \includegraphics[width=0.7\textwidth]{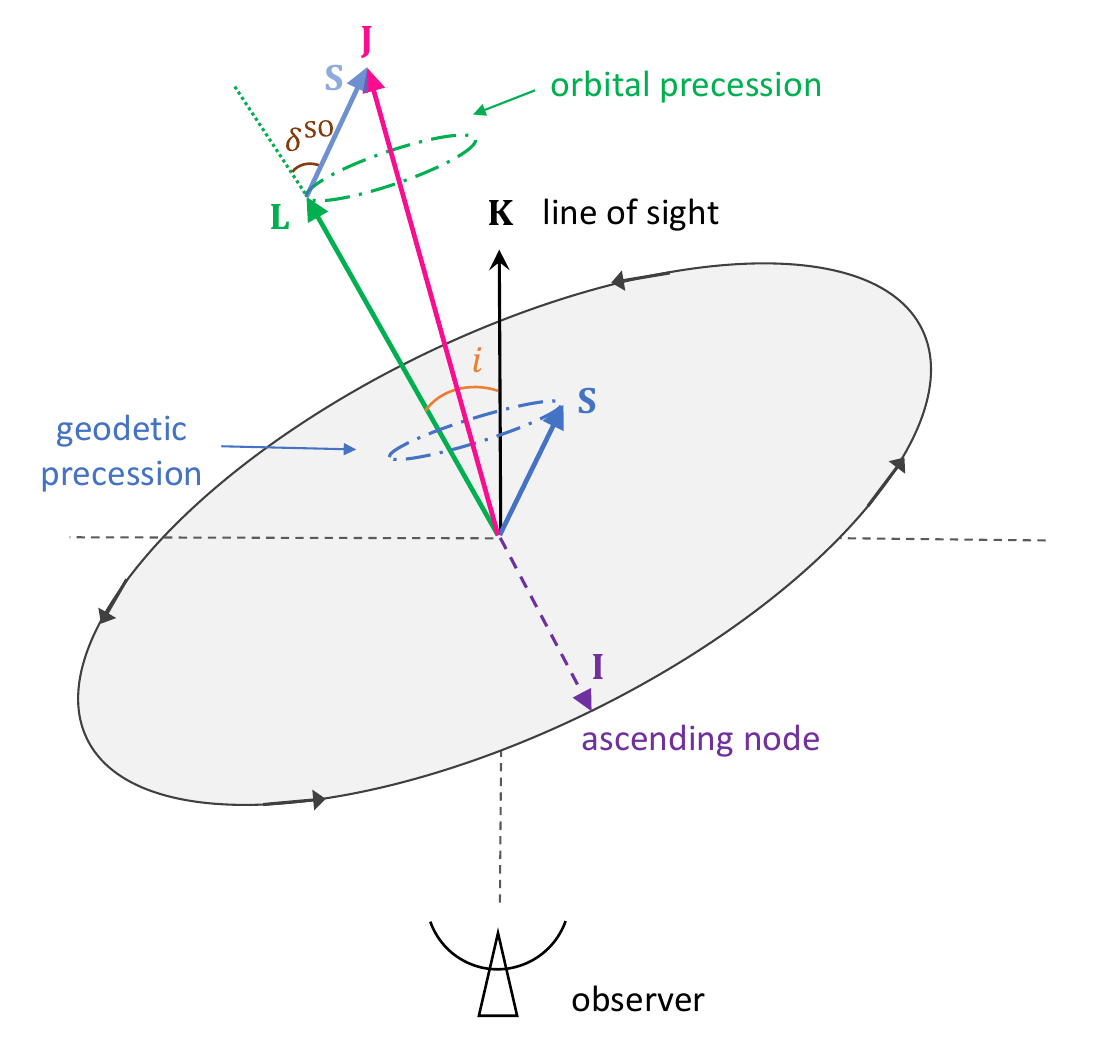}
    \caption{Orbital geometry of the system with all vectors shifted to the centre of mass of the system. \textbf{S} is the spin angular momentum of the pulsar (or companion) and \textbf{L} is the orbital angular momentum, which is perpendicular to the orbital plane and inclined at an angle $i$ to the line of sight vector, \textbf{K}. The total angular momentum vector $\textbf{J} = \textbf{L} + \textbf{S}$ and $\delta^{\rm SO}$ is the misalignment angle between \textbf{L} and \textbf{S}. As a result of this misalignment, both spin and orbit precess around \textbf{J}. The unit vector $\mathbf{I}$ points from the centre of mass to the ascending node.}
    \label{fig:orbit}
\end{figure}

Any vector undergoing parallel transport in a curved spacetime experiences a change in direction as seen from distant observers. If the spin axis of a pulsar is misaligned with the total angular momentum of the system $\textbf{J}$, it will slowly precess around $\textbf{J}$ at a rate of \cite{Damour1974,Barker1975PRD}
\begin{equation}
\Omega^\mathrm{geod} = \left(\frac{P_{\rm b}}{2 \pi} \right)^{-5/3}\, \frac{T_\odot^{2/3}}{1-e^2} \frac{m_\mathrm{c}(4m_\mathrm{p}+3m_\mathrm{c})}{2M^{\,4/3}}
\label{eq:Omega_geo}
\end{equation}
as predicted by General Relativity (GR). An illustration of the orbital geometry is shown in Fig.~\ref{fig:orbit}. 
$P_\mathrm{b}$ is the orbital period and the constant $T_\odot \equiv (\mathcal{GM})_\odot^\mathrm{N}/c^3$, with $(\mathcal{GM})_\odot^\mathrm{N} \equiv 1.327\,124\,4\times10^{26}\,  \mathrm{cm^3\,s^{-2}}$ being the nominal solar mass parameter defined by the IAU 2015 Resolution B3 \citep{Prsa2016} and $c$ the speed of light. 
$m_\mathrm{p}$ and $m_\mathrm{c}$ are the masses of the pulsar and the companion, and $M= m_\mathrm{p}+ m_\mathrm{c}$ is the total mass of the system.
In this and all following equations with $T_{\odot}$, the masses are dimensionless, being specified in nominal solar masses by taking the ratio $G M_\mathrm{object} / (\mathcal{GM})_\odot^\mathrm{N}$, where $G$ is the gravitational constant, and the nominal solar mass is defined as $\rm M_{\odot} = (\mathcal{GM})_\odot^\mathrm{N} /G$. 
The time eccentricity $e$ corresponds to the observed eccentricity included in the timing model \cite{DD86}. 

As the angle between the pulsar's spin and our line of sight varies with time, different parts of the emitted beam are observed and a temporal evolution of the pulse shape and polarisation is expected \cite{Barker1975ApJ,Esposito1975}. The first evidence of this relativistic spin precession was provided by the narrowing of the pulse profile observed in the Hulse-Taylor pulsar B1913+16 \cite{Kramer1998}. The full geometry of the system was determined by fitting a cone-like beam model and the misalignment of the pulsar spin to the orbital momentum vector was measured. Later, Weisberg \& Taylor confirmed these results using new high precision Arecibo data \cite{WT2002ApJ}.

Henceforth, the geodetic precession has been detected through profile variations in other recycled pulsars, such as PSR~B1534+12 \cite{Stairs2004}, PSR~B2127+11C \cite{Ridolfi2017PhDT}, and more recently in PSR~J1757$-$1854 \cite{Cameron2023} and PSR~J1946+2052 \cite{Meng+2024}. As we will discuss in more detail below, in these systems the orbital plane is no longer aligned with the spin axis of the recycled pulsar (thus creating the misalignment between the spin axis of the pulsar and $\textbf{J}$), the reason for this is the kick associated with the second supernova (SN) in the system.

In fact, geodetic precession has been discovered in all binary systems where the precession rate is expected to be sufficiently high (see \cite{Kramer2012mgm, Kramer2023mgm} for a review).
These also include pulsars that are the second-formed compact object in a binary system: PSR~J0737$-$3039B \cite{Burgay2005}, PSR~J1141$-$6545 \cite{Hotan2005} and PSR~J1906+0746 \cite{Lorimer_2006,Desvignes2019}. 
In particular, the precession rate of PSR~J0737$-$3039B has been measured from the observed changes of the eclipse light curve of PSR~J0737$-$3039A when B passes in front of it \cite{Breton2008,Lower+24}.
Indeed, for these pulsars, there is no special reason why their spin axes should be aligned with $\textbf{J}$, because they were never recycled by matter coming from the companion (which has angular momentum in the orbital plane). 
These younger pulsars provide the best measurements of $\Omega^\mathrm{geod}$ \cite{Desvignes2019, Lower+24} that match GR's prediction (Eq.~\ref{eq:Omega_geo}) within measurement uncertainties.
\\

%%%%%%%%%%%%%%%%%%%%%%%%%%%%%%%%%%%%%%%%%%
\section{The Lense-Thirring precession}
\label{sec2}

In GR, the gravitational field of a body has contributions from the mass currents associated with the body's proper rotation. The rotation of a massive body, such as the Sun, causes a dragging on the spacetime and hence makes the orbits of nearby masses (i.e. planets) precess. According to a 2007 historical study by Pfister \cite{Pfister_2007}, such dragging effect was first calculated by Albert Einstein in 1913 in the so-called Einstein-Besso manuscript (see pp. 344-473 in \cite{Klein}) for the tensorial Entwurf theory \cite{Entwurf}. In 1918, Josef Lense and Hans Thirring, with substantial help from Einstein, calculated the frame-dragging effect of rotating central bodies on the motion of planets and moons in GR \cite{Thirring1918, LT1918, Mashhoon+1984}, which is nowadays well known as the ``Lense-Thirring effect'' or ``Lense-Thirring precession''. 

Experimental tests of the LT precession were first proposed in the 1970s by launching two counter-orbiting drag-free satellites in polar orbit around the Earth \cite{VanPatten1976, VanPatten1976b}. 
Later, after the LAGEOS (LAser GEOdynamcs Satellite of NASA) being launched in 1976, Ciufolini proposed a new method of measuring the LT effect using a second LAGEOS satellite with an opposite orbital orientation \cite{Ciufolini1986}. 
The LAGEOS 2 was then launched by the Italian Space Agency (ASI) and NASA in 1992, followed by another laser-ranged satellite of ASI launched in 2012, the LARES (LAser RElativity Satellite) \cite{LARES2010}. Based on data collected with these three satellites, the LT precession test in Earth’s gravity field has reached an accuracy of about 2\% \cite{Ciufolini_2004, Ciufolini_2019}.

Another manifestation of the frame-dragging, which should not be confused with the LT precession, is the ``Schiff precession", where, on top of the effect of the geodetic precession on the spin of a test mass, there is a secondary effect caused by the spin of the massive body  \cite{Schiff1960PRL}. This resulted in the proposal for Gravity Probe B, which was eventually launched and measured both the geodetic and Schiff precessions, the latter matches the GR prediction well within its 19\% relative uncertainty \cite{GP-B}. This is the first measurement of ``spin-spin" coupling.

Similar to the weak-field scenario (i.e., the Solar System), the spins of compact objects (e.g. pulsars) in relativistic binaries are gravitationally coupled to the orbital motion of the systems \cite{Barker1975PRD}.  
This relativistic spin-orbit coupling leads to an additional contribution to the rate of change of periastron $\dot{\omega}$ and a precession of the orbital plane about the direction of the system's total angular momentum (see Fig.~\ref{fig:orbit}). In the latter case, it induces a change in the inclination angle $i$.   
In the following sections, we recapitulate these two cases, highlighting the detailed predictions for the motions for binary pulsars, as these are the only ones where we might detect LT precession caused by the rotations of NSs.

%%%
\subsection{Spin-orbit coupling in periastron advance}
\label{subsec:SO}
The spin-orbit coupling yields a contribution to the advance of periastron, $\dot{\omega}$. According to \cite{DS88}, in GR the total observed rate of periastron advance $\dot{\omega}^\mathrm{obs}$, up to the second post-Newtonian (PN) order, can be written as
\begin{align}
\dot{\omega}^\mathrm{obs} = \frac{2\pi}{P_\mathrm{b}} \,k\,,
\end{align}
with the dimensionless periastron advance parameter $k$ written in the form \cite{DS88,Kramer+2021DP}:
\begin{align}
k
&= \frac{3\,\beta_\mathrm{O}^2}{1-e^2} \left(1 + f_\mathrm{O}^{}\, \beta_\mathrm{O}^2 - g_\mathrm{S_A}^{}\, \beta_\mathrm{O}^{}\, \beta_\mathrm{S_A}^{} - g_\mathrm{S_B}^{}\, \beta_\mathrm{O}^{}\, \beta_\mathrm{S_B}^{} \right) \,,
\end{align}
where
\begin{align}
&\beta_\mathrm{O}^{} = \left(\frac{2\pi}{P_\mathrm{b}} T_\odot M \right)^{1/3} \,,\\
&\beta_{\mathrm{S}_j}^{} = \frac{G\, S_{j}}{c^5\, T_\odot^2\, m_j^2}  
=\frac{G \,I_j\,2 \pi \nu_j}{c^5\, T_\odot^2\, m_j^2} \,,\\
&f_\mathrm{O}^{} = \frac{1}{1-e^2} \left(\frac{39}{4}X_\mathrm{A}^2 + \frac{27}{4}X_\mathrm{B}^2 + 15 X_\mathrm{A}^{} X_\mathrm{B}^{} \right) -\left(\frac{13}{4}X_\mathrm{A}^2 + \frac{1}{4}X_\mathrm{B}^2 + \frac{13}{3} X_\mathrm{A}^{} X_\mathrm{B}^{} \right),\\
&g_{\mathrm{S}_j}^{} = \frac{X_j\,(4X_j+3X_{j'})}{6\,(1-e^2)^{1/2} \sin^2 i} \left[( 3\sin^2 i -1)\,\textbf{k}\cdot\textbf{s}_j +\cos(i) \,\textbf{K} \cdot \textbf{s}_j\right] \,.\label{eq:g_S}
\end{align}
where $e$ is orbital eccentricity, $n_\mathrm{b} = 2\pi/P_\mathrm{b}$ is the orbital frequency with $P_\mathrm{b}$ being the orbital period.
The subscript A and B stands for the two bodies in the binary (in the case of the Double Pulsar, i.e. pulsar A and pulsar B). 
The terms with subscript O are associated with orbital contributions and with S is associated with spin contributions. 
The subscript $j$ denotes one body ($j$ = A, B) and $j'$ corresponds to the other body of the system. $S_j$ is the magnitude of the spin of body $j$, which equals the product of the MoI $I_j$ with $2 \pi \nu_j$, where $\nu_j$ is the spin frequency. 
The mass ratio of one mass component, $m_j$, and the total mass of the system, $M = m_\mathrm{A} + m_\mathrm{B}$, is indicated by $X_j = m_j/M$. $\textbf{s}_j$ is the unit spin vector of body $j$, $\textbf{k}$ the unit vector in the direction of the orbital angular moment, and $\textbf{K}$ the unit vector along the line of sight. 

%%%%%%%%%%%%%%%%%%%%%%%%%%%%%%%%%
\subsection{Spin-orbit coupling in orbital inclination}
\label{subsec:inc}

Another way of measuring LT precession is through the changes in the orbital inclination angle $i$ and subsequently in an observable, the projected semi-major axis of the pulsar orbit $x = a_\mathrm{p} \sin{i} /c$ ($a_\mathrm{p}$ is the semi-major axis of the pulsar orbit). This leads to a precession in the projected semi-major axis \cite{Barker1975PRD}
\begin{equation}
    \dot{x}^\mathrm{LT} = x \cot{i}\, \left(\frac{\mathrm{d}i}{\mathrm{d}t}\right)^\mathrm{LT} \,, 
    \label{eq:xdot_LT}
\end{equation}
For nearly edge-on systems ($i \approx 90^{\circ}$), this contribution is small as $\cot{i}\ll 1$ and mostly likely cannot be detected, independently of the cause of $\mathrm{d}i/\mathrm{d}t$.

The $(\mathrm{d}i / \mathrm{d}t)^\mathrm{LT}$ term is given, in GR, by:
\begin{equation}
    \left(\frac{\mathrm{d}i}{\mathrm{d}t}\right)^\mathrm{LT} = \frac{3}{2}\,
    \left( \frac{P_{\mathrm{b}}}{2\pi} \right)^{-2}
     \frac{G}{c^5 T_{\odot} M (1-e^2)^{3/2}}
     \sum_{j}\, \left(\frac{1}{X_j} + \frac{1}{3} \right) \,(\mathbf{S}_j \cdot \mathbf{I})\,,
\end{equation}
where $\mathbf{S}_j$ is the spin angular momentum (``spin'') of body $j$,  and $\mathbf{I}$ is a unit vector pointing from the centre of mass of the system to the ascending node.

In binary pulsar systems, the total observed change in the projected semi-major axis of the pulsar orbit can be written as \cite{LK2004}:
\begin{equation}
    \dot{x}^\mathrm{obs} = \dot{x}^\mathrm{PM} + \dot{x}^\mathrm{\dot{D}} + \dot{x}^\mathrm{GW} + \dot{x}^\mathrm{\dot{m}} + \dot{x}^\mathrm{3rd} + \dot{x}^{\dot{\epsilon}_\mathrm{A}} + \dot{x}^\mathrm{LT} \,,\label{eq:xdot}
\end{equation}
where the other contributions come from the proper motion of the system ($\dot{x}^\mathrm{PM}$), the changing radial Doppler shift ($\dot{x}^\mathrm{\dot{D}}$), GW emission ($\dot{x}^\mathrm{GW}$), mass loss in the system ($\dot{x}^\mathrm{\dot{m}}$), the presence of a hypothetical third body in the system $\dot{x}^\mathrm{3rd}$, and a secular change in the aberration of the pulsar beam due to geodetic precession $\dot{x}^{\dot{\epsilon}_\mathrm{A}}$, respectively.

%%%%%%%%%%%%%%%%%%%%%%%%%%%%%%%%%%%%%%%%%%
\section{Evolution of binary pulsars}
\label{sec:evo}

\begin{figure}[t]
    \centering
    \includegraphics[width=0.85\textwidth]{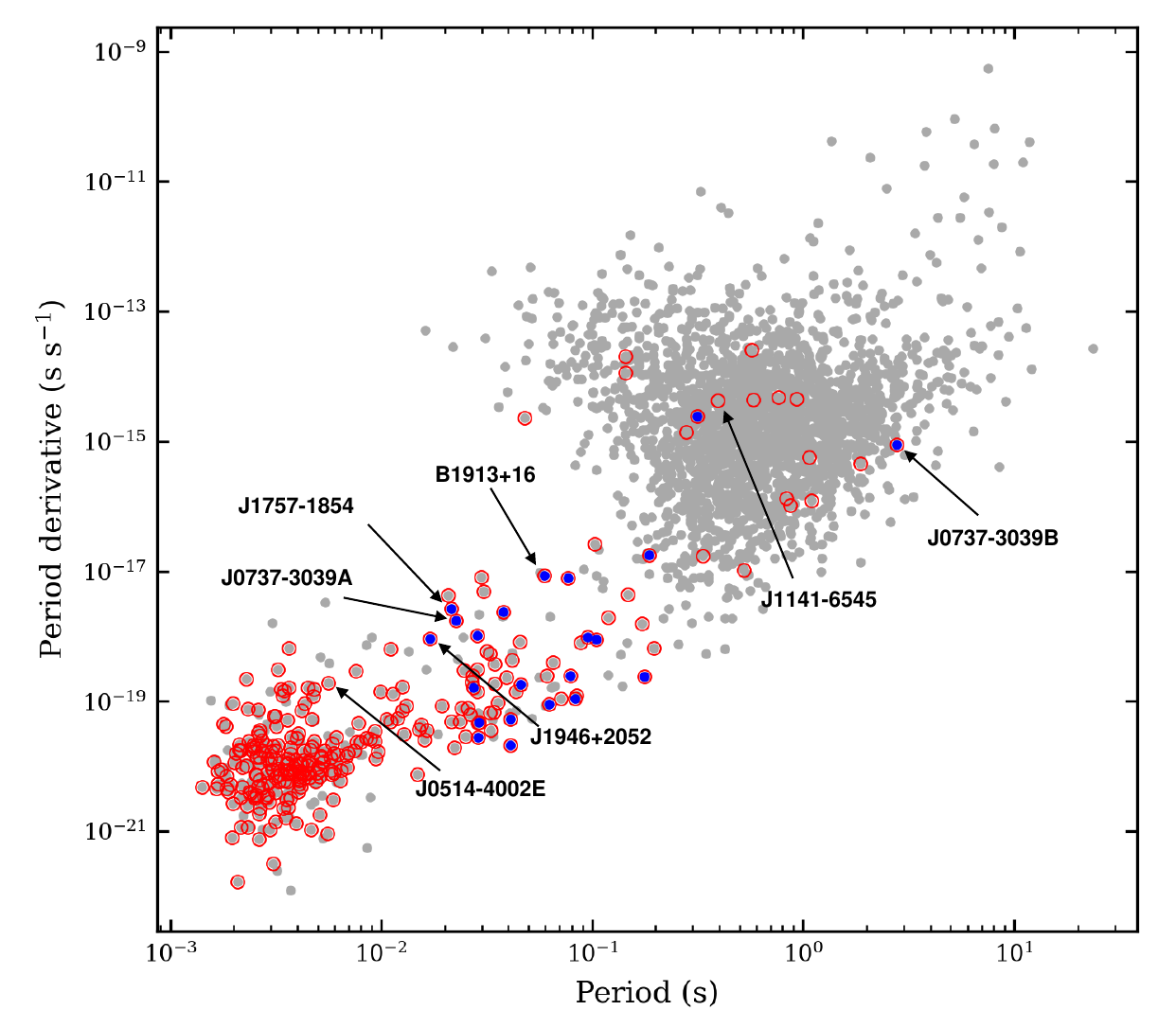}
    \caption{$P-\dot{P}$ diagram of known pulsars plotted in log-log scale. Pulsars on the upper right with spin period $\sim 10^{-1}-10^1$~s are known as normal pulsars, whereas pulsars on the bottom left with spin period of milliseconds are called millisecond pulsars (MSPs). Pulsars in binary systems are marked with a red circle, and  the DNS systems are highlighted in blue. Binary pulsars detailed in this article are marked by black arrows with their name labelled. Data were taken from the ATNF Pulsar Catalogue version 2.0 \cite{Manchester+2005}. }
    \label{fig:ppdot}
\end{figure}

In Fig.~\ref{fig:ppdot}, we plot the spin period ($P$, horizontal axis) and the spin period derivative ($\dot{P}$, vertical axis) for most known radio pulsars (grey dots). These appear in two main groups: the more numerous, which we will call the "normal" pulsars, form a large clump around $P \, = \, 1$ s and $\dot{P} = 10^{-15} \rm \, s \, s^{-1}$. The second group appears in the lower left, with significantly smaller values of $P$ and $\dot{P}$, and are known as the "recycled" pulsars.

The latter group are known as "recycled" because, unlike the normal pulsars, they were spun up (and their magnetic fields degraded) by accretion of mass from a stellar companion after they formed and died. 
The dots with red circles indicate the pulsars that are in binary systems: in the lower left corner, about 80\% of pulsars are in binaries, while very few "normal" pulsars are in binaries.

\subsection{Types of binary pulsars}

\begin{figure}[t]
    \centering
    \includegraphics[width=\textwidth]{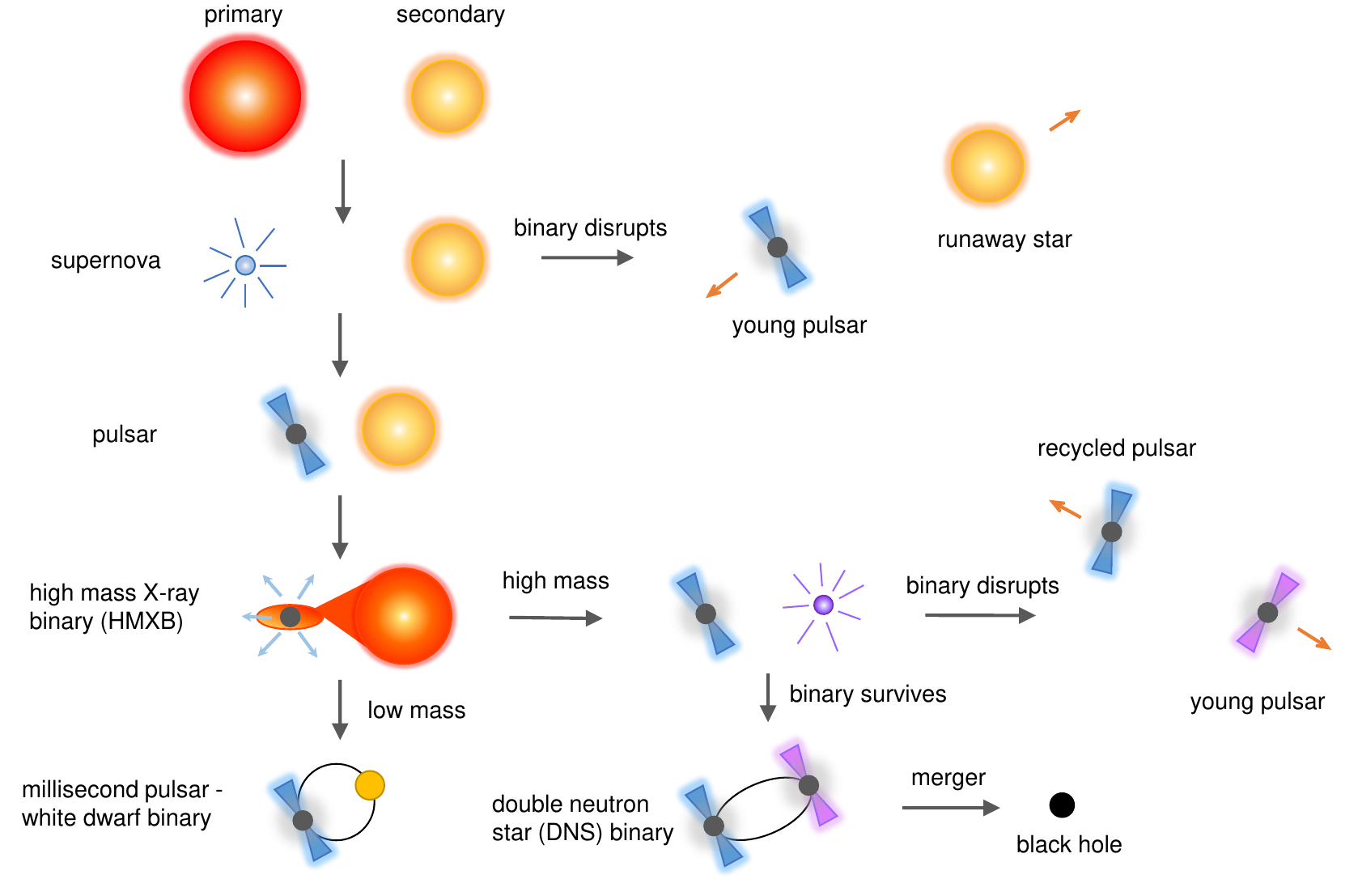}
    \caption{Schematic diagram showing the evolution scenario of various pulsar systems, based on the concepts in \cite{Lorimer2001}.}
    \label{fig:evolution}
\end{figure}

In Fig.~\ref{fig:evolution}, we show in a very schematic way the two main formation channels for binary pulsars (for a detailed review, see \cite{tauris23}). For simplicity, we start with a binary system consisting of two main sequence stars. This is a common occurrence, especially for massive stars, the majority of which form in binary and multiple systems \cite{Sana2012}.

As the more massive star (the primary) evolves, it eventually explodes as a SN, forming a "normal" pulsar. The system will then very likely disrupt owing to the kick and mass loss associated with this SN; we know this not only from the aforementioned fact that most "normal" pulsars are "single", but also from the observation that surviving systems (consisting of a pulsar and a massive main sequence star) have very high orbital eccentricities (generally of order 0.9, e.g. PSR~B1259$-$63 with $e = 0.87$ \cite{Shannon_14}, PSR~J0045$-$7319 with $e = 0.808$ \cite{Kaspi_96}, PSR~J1638$-$4724 with $e = 0.955$ \cite{Lorimer_ParkesMB}, and PSR~J2032+4127 with $e = 0.96$ \cite{Ho_17}), indicating a near disruption.

Eventually, in some of these systems, the secondary will also evolve and become a giant star. Its large size will cause several effects, the first being tidal circularisation of the orbit. When the secondary fills its Roche lobe, the transfer of matter to the NS starts, and the system becomes an X-ray binary. This mass transfer can become unstable, leading to a common envelope phase. At this stage, the NS is slowly spun up, and its spin will become aligned with the orbital angular momentum. This process is though to ablate the magnetic field of the NS.

What happens next depends on the mass of the secondary:
\begin{itemize}
\item If it is massive enough, it will also go through a SN explosion, the system then either disrupts (a likely occurrence) or forms an eccentric DNS system, which will consist of a recycled pulsar and a short-lived "normal" pulsar. The high risk of disruption makes these systems relatively rare ($< 1\%$). 
These systems are necessarily eccentric because of the mass loss and the kick associated with the SN (the average kick for young isolated radio pulsars is $400-500 \, \mathrm{km\, s^{-1}}$ \cite{Lyne_Lorimer1994Natur, Hobbs+2005}, this is likely to be slightly less in binary systems). Because of this kick, the orbital plane after the SN might become very different from what it was before the explosion, causing a misalignment between the spin of the recycled pulsar and the orbital angular momentum. 
The most compact of these systems will coalesce in a Hubble time, forming a light black hole (BH, for an extensive discussion on these systems, see \cite{tauris17}).
\item If the secondary is light, then it will slowly evolve into a WD star. In this case the system will retain a low orbital eccentricity and the alignment between the spin and orbital angular momentum.
\end{itemize}

As we can see in Fig.~\ref{fig:ppdot}, the recycled pulsars with low-mass WD companions (commonly known as "millisecond pulsars", MSPs) have spin frequencies one order of magnitude larger than the recycled pulsars with NS companions. 
The reason is that for NSs with low-mass companions, they spend much longer (up to $\sim 10^8$~yr) as X-ray binaries because their evolution is much slower.
The resulting high spin frequencies give MSPs high timing precision.

The DNS systems have a known range of eccentricities from 0.06 to 0.83. These eccentricities are determined by a range of factors, like the masses of the components of the system before the second SN, the orbital period of that binary, and the mass loss and the magnitude and orientation of the kick that happen during the second SN.
The observations indicate that the dominant factor appears to be the mass of the second-formed NS: if it is smaller than $1.35 \, \rm M_{\odot}$, then $e < 0.3$; if it is larger, then $e > 0.55$ \cite{tauris17}. This suggests that SN kicks are linked to the mass of the newly formed NS.

Among the DNSs with lower-mass companions, these seems to be a hard floor for the eccentricities. Even for a symmetric SN explosion with negligible or small amount of ejecta mass, one would expect an eccentricity after the SN of  $e=\Delta M / (M_\mathrm{NS,1} + M_\mathrm{NS,2}) \simeq 0.06-0.13$, where $\Delta M$ is the amount of instantaneous mass loss from the exploding star \cite{tauris17}, which is the binding energy of the NS.

%%%%%%%%%%%%%%%%%%%%%%%%%
\subsection{Suitability for measurements of the LT precession}\label{sec:suit}

Despite the better timing precision of MSPs, and their much larger angular momenta (and thus potential LT effects), DNS systems are preferable laboratories for testing the LT effect. 
This is because the small eccentricities observed in the vast majority of MSP-WD systems ($10^{-7}$ to $10^{-4}$) make it impossible to measure $\dot{\omega}$ precisely and measure the LT contribution to that effect. Furthermore, as discussed above, the spin of the recycled pulsar and the orbital angular momentum remain aligned, resulting in a null value of $\dot{x}^{\rm LT}$. Therefore, MSP-WD systems are generally not suitable for detecting the LT precession.
An exception is PSR~J1141$-$6545, which has undergone a different evolutionary scenario resulting in a slow pulsar spin, a relatively high eccentricity and a fast spin of the massive WD. This is  discussed in more detail in Section~\ref{sec:J1141}.

However, some fast-spinning MSPs might still be useful for these purposes. Because of the very high stellar densities of globular clusters, a MSP in that environment can exchange its low-mass WD companion by a much more massive companion (see e.g., \cite{Prince+91,Freire+2004}). The resulting orbits will generally be very eccentric and misaligned with the spin of the MSP. This and the high precision timing of MSPs could make such systems promising for measuring the relatively large $\dot{x}^\mathrm{LT}$ that results from their large spins, especially if the systems are compact. Things become even more interesting if the massive companion is a fast-spinning BH --- a possibility in the case of NGC 1851E binary MSP, discussed in Section~\ref{sec:J0514} --- then the LT precession would be a promising way of measuring the spin of the BH companion.

Finally, we note that the aforementioned binary pulsars with a main sequence star companion are inadequate for this sort of measurements. First, because they are slowly spinning normal pulsars, their timing precision is generally much worse than recycled pulsars. Second, those systems are also much wider, making the detection of relativistic spin-orbit coupling impossible even with high precision timing. Third, any relativistic spin-orbit coupling would originate from the much larger angular momentum of the main sequence star. Finally, in those systems the classical spin-orbit coupling and tidal effects are many orders of magnitude larger than the relativistic spin-orbit coupling \cite{wex98}, and have been detected in the timing \cite{Lai+1995ApJ,Shannon_14}.

%%%%%%%%%%%%%%%%%%%%%
\section{Measuring the LT precession via pulsar timing}
\label{sec:psrs}

The first binary pulsar where the presence of the LT precession was shown was PSR~J1141$-$6545 \cite{Vivek+2020Sci}. In this case, the precession is not caused by the pulsar, but instead by a fast-spinning, massive WD companion, and is superposed (to an unknown extent) to classical spin-orbit coupling caused by the quadrupolar moment of the WD. This system shows some of the possibilities and limitations of pulsar timing measurements of the LT precession.

The promise of the first measurement of the MoI of a NS arrived with the discovery of the Double Pulsar system PSR J0737$-$3039A/B in 2003, because of its extremely compact 2.45-h orbit \citep{Burgay2003,Lyne2004} and the alignment of the spin of pulsar A with the orbit, which simplifies the analysis.

\begin{figure}[t]
    \centering
    \includegraphics[width=0.85\textwidth]{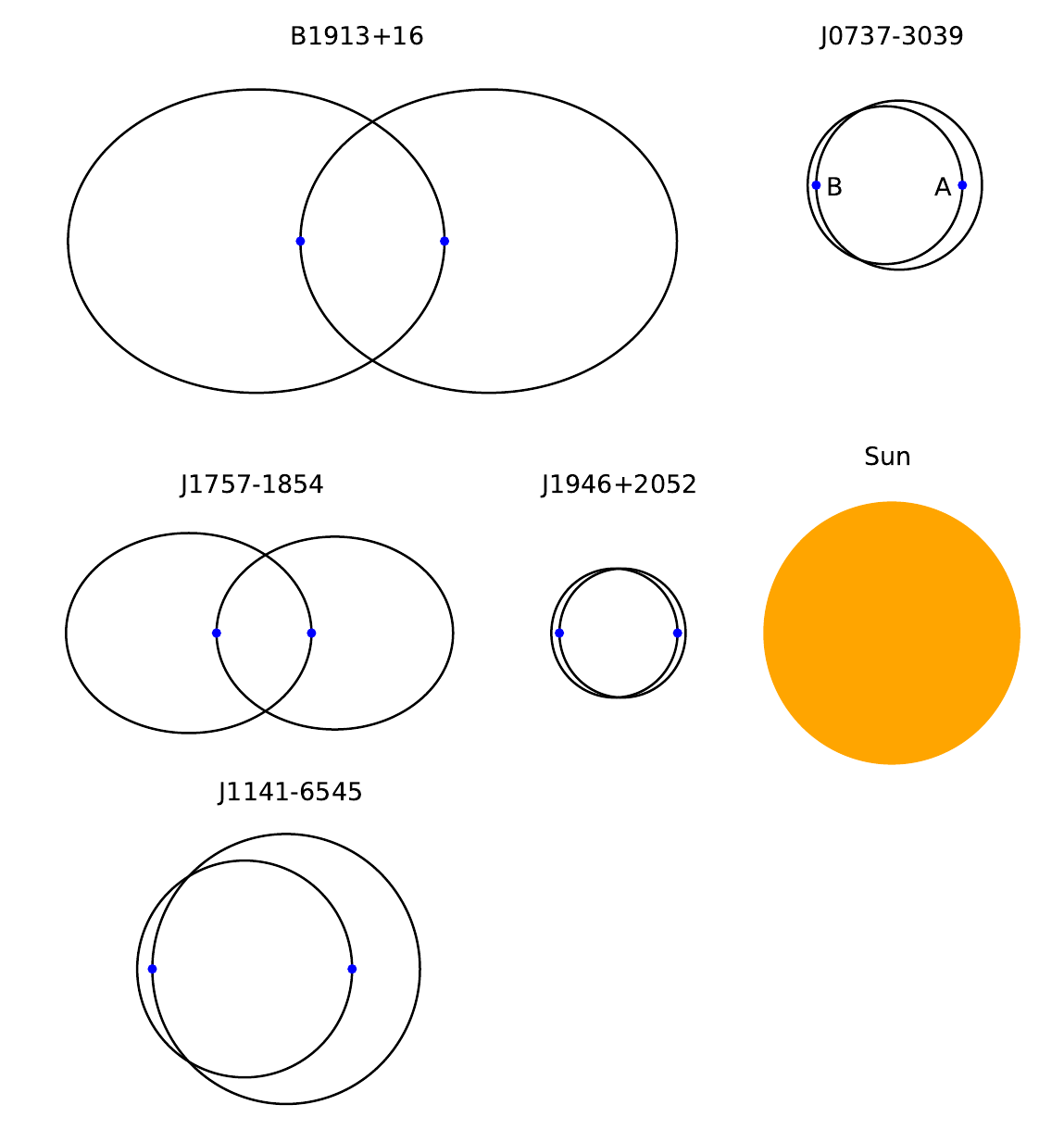}
    \caption{Schematic diagram of the compact binary pulsar systems discussed in the text (B1913+16 \cite{Weisberg+2016ApJ}, J0737$-$3039 \cite{Hu2022}, J1757$-$1854 \cite{Cameron2023}, J1946+2052 \cite{Stovall+2018ApJ}, and J1141$-$6545 \cite{Vivek+2020Sci}) compared to the size of the Sun. The left orbit of each pair corresponds to the pulsar whereas the right orbit corresponds to the companion (pulsar A and pulsar B in the case of the Double Pulsar). The size of NSs are scaled  $\sim 10^3$ times larger for illustration purpose.}
    \label{fig:dns}
\end{figure}

In the following we discuss these two systems plus two additional binaries where the measurement of the LT effect appears to be promising at the current time: PSR~J1757$-$1854 and PSR~J1946$+$2052. A schematic comparison of these compact binary systems with the the Hulse--Taylor pulsar PSR~B1913$+$16 and the Sun is shown in Figure~\ref{fig:dns}.

%%%%%%%%%%%%%%%%%%%%%%%%%%%%%%%
\subsection{Lense-Thirring precession in \texorpdfstring{$\dot{x}$}{Lg}}
\subsubsection{PSR~J1141\texorpdfstring{{$-$}}{Lg}6545, a young pulsar with a massive white dwarf companion}
\label{sec:J1141}

PSR J1141$-$6545 is an atypical system among pulsar-WD systems, in the sense that the pulsar is not recycled --- its spin period of 396~ms and large spin period derivative place it firmly among the ``normal'' pulsars (see Fig.~\ref{fig:ppdot}). Furthermore, unlike most other pulsar-WD systems, its orbit has a significant eccentricity ($e=0.17$), and its spin axis is misaligned with the orbital angular momentum, as deduced from the variation of its pulse profile  \cite{Hotan2005}. That the companion is a massive WD was established optically \cite{antoniadis2011}.

Thus, the system displays all the characteristics typical of DNS systems, in particular the characteristics of the pulsar are similar to those of PSR J0737$-$3039B, the second-formed pulsar in the ``Double Pulsar'' system (see \ref{sec:0737} for details). Why is the primary, then, a WD? The explanation was published in 2000 by Tauris \& Sennels \cite{tauris2000}. At the start of the evolution of the system, the primary lost so much mass to the secondary during its evolution that it went under the mass limit for the formation of a NS, forming instead a massive WD. The secondary, after accreting all the mass from the primary, becomes massive enough to undergo a SN, forming then a NS, now visible as a normal pulsar.

A consequence of this evolution is that, before the SN explosion, the progenitor to the pulsar fills its Roche lobe and starts transferring matter to the WD, thus spinning it up and aligning its angular momentum with the orbital angular momentum. The mass and kick associated with the subsequent SN made the system eccentric and changed the orbital plane, as in a DNS system. Since then the direction of the spin of the WD has been, like that of the pulsar, precessing around the total angular momentum via geodetic precession.
The orbital precession is in this case overwhelmingly due to the spin of the WD, not the NS, which has a MoI $\sim 10^5$ times smaller than the WD, but only spins about $\sim 10^2$ times faster.

\begin{figure}[t]
    \centering
    \includegraphics[width=0.85\textwidth]{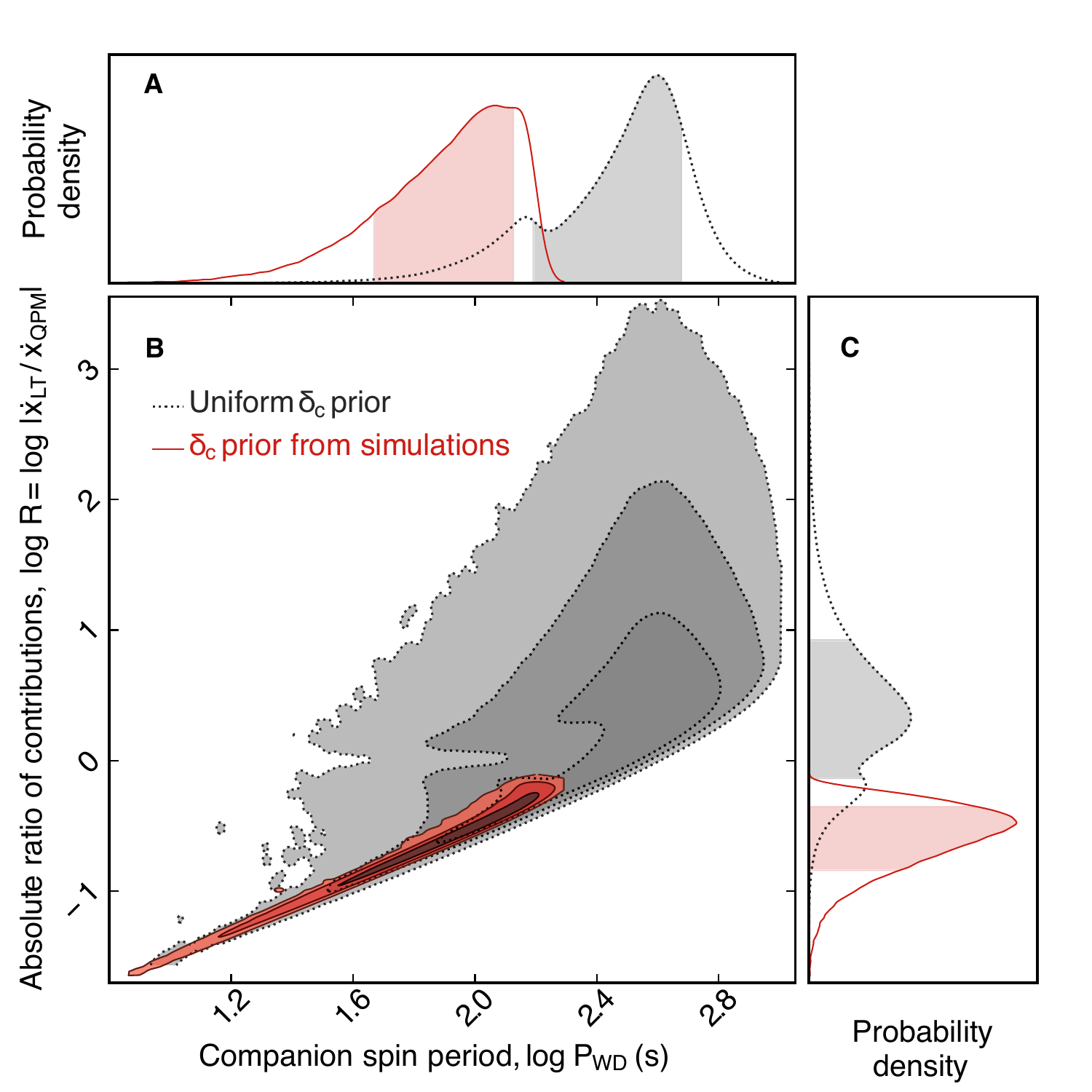}
    \caption{The absolute ratio of the contributions to $\dot{x}^{\rm SO}$ from $\dot{x}^{\rm LT}$ and $\dot{x}^{\rm QPM}$, $R = | \dot{x}^{\rm LT} / \dot{x}^{\rm QPM}|$, plotted as a function of the unknown spin period of the WD ($P_{\rm WD}$). (A and C) Marginalised posterior distributions with their 68\% confidence intervals shaded, defined as the combination of the two 34\% confidence regions on either side of the 2D maximum of the likelihood function. (B) Two-dimensional probability distribution with contours defining the 68\%, 95\%, and 99\% likelihood confidence intervals. The gray-shaded regions and dotted contours are constraints using only the radio observations of the pulsar, whereas the red regions and solid contours include additional binary evolutionary constraints from simulations. Figure from \cite{Vivek+2020Sci}. Reprinted with permission from AAAS.}
    \label{fig:1141}
\end{figure}

The large MoI of the fast-spinning WD explains why it was in this pulsar, which has a relatively low timing precision, that the spin-orbit coupling was seen \cite{Vivek+2020Sci}. The effect was detected as a variation of the projected semi-major axis ($\dot{x}$) caused by the variation of the inclination angle of about 14 arcseconds per year. The effect could only be detected by assuming that GR is the correct gravity theory, so it cannot be interpreted as a test of the theory. Unlike in the case of NSs, where the $\dot{x}$ caused by spin-orbit coupling ($\dot{x}^{\rm SO}$) comes entirely from the LT precession ($\dot{x}^{\rm LT}$, see Eq.~\ref{eq:xdot_LT}), and unlike in main sequence stars, where this effect would arise from classical spin-orbit coupling due to the quadrupolar moment of the star ($\dot{x}^{\rm QPM}$), in massive, fast-spinning WDs both contributions to $\dot{x}^{\rm SO}$ are of a similar order of magnitude.

What percentage of the $\dot{x}^{\rm SO}$ arises from $\dot{x}^{\rm LT}$ and $\dot{x}^{\rm QPM}$ depends on the unknown spin period of the WD ($P_{\rm WD}$), the unknown misalignment angle between the spin of the WD and the orbital plane ($\delta_\mathrm{c}^{\rm SO}$) and a precession phase $\Phi_\mathrm{c}$, see Fig.~\ref{fig:1141}. Again, these contributions were calculated strictly within the framework of GR, which was used to detect $\dot{x}^{\rm SO}$ in the first place. 

Fig.~\ref{fig:1141} highlights some of the limitations of measurements of the LT effect based on $\dot{x}$, which is compounded in this case by the fact that the spin period of the companion WD is unknown. In a DNS system, we have a few advantages: a) For NSs, $\dot{x}^{\rm LT}$ is several orders of magnitude larger than $\dot{x}^{\rm QPM}$, which means that the latter term can be ignored, b) we generally know the spin period of the recycled pulsar\footnote{The misaligment of the spin of a second-formed NS causes a much smaller effect on $\dot{x}^{\rm LT}$, because such pulsars are generally much slower than any recycled pulsars. For this reason, only the misalignment of the first-formed, recycled object matters.}.
Nevertheless, even in that case, lack of knowledge of $\delta_\mathrm{c}^{\rm SO}$ and $\Phi_\mathrm{c}$ will still limit the measurement of the angular momentum of the NS and its MoI from a measurement of $\dot{x}$. 

It is important to emphasise that, for Fig.~\ref{fig:1141}, the MoI of the WD is entered as a given, {\em not} as a quantity we are trying to measure. The reason for this is threefold: first, unlike in the case of NSs, the MoI of WDs is reasonably well known; second, it tells us nothing about the state of super-dense nuclear matter, which does not exist in WDs; third, even if we tried to measure it, the fact that there is only one measurement ($\dot{x}^{\rm SO}$) and three unknowns ($P_{\rm WD}$, $\delta_\mathrm{c}^{\rm SO}$ and $\Phi_\mathrm{c}$) makes that measurement impractical.

So, what do we learn from Fig.~\ref{fig:1141}? Two things: a) the slower the spin of the WD, the larger is $\dot{x}^{\rm LT}$ relative to $\dot{x}^{\rm QPM}$, the reason is that the former varies with $P_{\rm WD}^{-1}$ and the latter with $P_{\rm WD}^{-2}$, b) we can indeed constrain $P_{\rm WD}$ to a range from 10 to $\sim$150 s, if we take into account the configurations that emerge from a simulation that takes into account the system's orbital parameters (orange contours). This is a very fast spin period for a WD, and an unexpected confirmation of the evolutionary scenario described by Tauris \& Sennels \cite{tauris2000}. 

\subsubsection{PSR~J1757\texorpdfstring{$-$}{Lg}1854, the most accelerated binary pulsar}
\label{sec:1757}

PSR~J1757$-$1854 is a 21.5-ms pulsar in a highly eccentric ($e=0.61$) 4.4-hr orbit around a NS companion. This pulsar was discovered in 2016 with the 64-m Parkes radio telescope Murriyang as part of the High Time Resolution Universe Pulsar Survey \cite{Cameron2018}. The orbit of this system resembles a compact version of the Hulse-Taylor pulsar PSR~B1913+16, with one of the shortest merger times of any known DNS ($\sim 76$~Myr) and highest relative velocity ($1060 \,\mathrm{km\,s^{-1}}$) and acceleration ($\sim 680\,\mathrm{m\,s^{-2}}$) at periastron \cite{Cameron2023}. 
This system was believed to be promising for measuring the LT precession to $\dot{x}$. As shown in Section~\ref{subsec:inc}, this contribution depends on the misalignment angle and geodetic precession phase. In the recent study \cite{Cameron2023}, the preferred geometry solutions obtained from ``precessional Rotating Vector Model'' analysis \cite{KW2009} indicate an upper limit for $\dot{x}^\mathrm{LT}$ (see Table~\ref{tab:LT}) two orders of magnitude smaller than the uncertainty of $\dot{x}^\mathrm{obs}$ from 6-yr campaign with the Green Bank Telescope (GBT) and Parkes telescopes. With simulated observations for GBT and MeerKAT, it was concluded that a 3-$\sigma$ measurement of $\dot{x}^\mathrm{obs}$ may be possible by 2031, and other contributing terms in Eq.~\eqref{eq:xdot} also need to be determined and deducted so as to measure the LT contribution, which is a challenging task.
{\renewcommand{\arraystretch}{1.3}
\begin{table}[h] 
\caption{The LT contribution to the rate of periastron advance in four aforementioned promising pulsar systems. \label{tab:LT}}
\centering
\begin{threeparttable}
\begin{tabularx}{0.9\textwidth}{lcc}
\toprule
\textbf{Pulsar name}	& $\quad\dot{\omega}^\mathrm{LT}$ (deg $\mathrm{yr}^{-1}$)	& $\quad\quad\dot{x}^\mathrm{LT}$ (lt-s $\mathrm{s^{-1}}$)\\
\midrule
PSR~J0737$-$3039A & $\quad\quad \quad-0.000377 \,\times\, I_\mathrm{A}^{(45)}$ $^{*}$ ~\cite{Hu+2020} & -\\
PSR~J1946+2052$^{**}$	  & $\,-0.000677 \,\times\, I_\mathrm{A}^{(45)}$  & -\\
PSR~J1757$-$1854 & - & $\quad |\dot{x}^\mathrm{LT}|\leq 2.01 \times 10^{-15} \times I_\mathrm{A}^{(45)}$ \cite{Cameron2023}\\
PSR~J0514$-$4002E & - & 
$\dot{x}^\mathrm{LT} \lesssim 1.7 \times 10^{-13}$ \cite{Barr2024} \\
\bottomrule
\end{tabularx}
\begin{tablenotes}
{\small 
$^{*}$ $I_\mathrm{A}^{(45)}=I_\mathrm{A} / (10^{45}\mathrm{g\,cm^2})$. } \\
$^{**}$  Mass is assumed to be $1.25\,\mathrm{M_\odot}$ for both NSs.
\end{tablenotes}
\end{threeparttable}
\end{table}
}
\subsection{Lense-Thirring precession in \texorpdfstring{$\dot{\omega}$}{Lg}}
\subsubsection{PSR~J0737\texorpdfstring{$-$}{Lg}3039A/B, the unique Double Pulsar system}
\label{sec:0737}

The Double Pulsar system PSR~J0737$-$3039A/B was discovered in 2003 \cite{Burgay2003}, with a pulsar in a compact orbit with a NS, which was also confirmed to be a pulsar shortly afterwards \cite{Lyne2004}. The system is composed of a 23-ms recycled pulsar ``A'' and a 2.8-s young pulsar ``B'' in a highly relativistic 2.45-h orbit. This remains the only known system comprising two pulsars, although pulsar B disappeared from our view since March 2008 due to geodetic precession \cite{Perera2010}.

The low orbital eccentricity ($e=0.088$), small transverse velocity of the system ($\sim 10 \,\mathrm{km \, s^{-1}}$) \cite{Kramer2006} and small misalignment angle between A's spin vector $\textbf{s}_\mathrm{A}$ and the orbital angular momentum vector $\textbf{k}$ ($\delta_\mathrm{A}^{\rm SO}<3.2\si{\degree}$) \cite{Manchester_2005,Ferdman2008,Ferdman2013} all support the opinion that the second-born pulsar ``B'' is formed in a low-kick SN event \cite{Willems2006,Stairs2006,tauris17}. Moreover, the rotation of A has been confirmed to be prograde using the modulation of B's emission by the interaction with A's pulsar wind \cite{Pol2018}. The unique high inclination angle between the orbital angular momentum and the line of sight towards the pulsar system ($i=89.36\si{\degree}\pm0.03\si{\degree}$ or $180\si{\degree}-i$ from timing analysis) enabled the studies of aberrational light bending effect \cite{Kramer+2021DP, Hu2022}, thus independently confirming the prograde rotation of A and providing additional evidence for a low-kick SN.

%%%%%%%%%%%%%%%%%%%%%%%%
\subsubsection{Measure the neutron star moment of inertia with \texorpdfstring{$\dot{\omega}, s$}{Lg} and \texorpdfstring{$\dot{P}_\mathrm{b}$}{Lg}}
\label{sec:MoI}

Since PSR~J0737$-$3039A/B is almost edge on to our line of sight, the LT contribution in $\dot{x}$ is insignificant. Therefore, the only practicable way to access the MoI of pulsar A, $I_\mathrm{A}$, is through the LT contribution in $\dot{\omega}$. 
Given that the spin of A is practically aligned with the orbital angular momentum, assuming $\textbf{s}_\mathrm{A} = \textbf{k}$, for pulsar A the expression of the spin contribution in Eq.~\eqref{eq:g_S} can be simplified to \citep{DS88}
\begin{align}
g_\mathrm{S_A}^\parallel = \frac{1}{(1 - e^2)^{1/2}} \left(\frac{1}{3} X_\mathrm{A}^2 + X_\mathrm{A}^{} \right)\,.
\end{align}
As pulsar B spins about 122 times slower than pulsar A, the LT contribution of pulsar B is negligible \cite{Hu+2020}.
Therefore, the total observed rate of periastron advance in the Double Pulsar system can be written as
\begin{align}
&\dot{\omega}^\mathrm{obs} = \dot{\omega}^\text{1PN} +\dot{\omega}^\text{2PN} +\dot{\omega}^\text{LT}_\mathrm{A} \,,\label{eq:omdot}
\end{align}
where the 1PN and higher-order corrections due to 2PN effects and LT contribution of pulsar A are 
\begin{align}
\dot{\omega}^\text{1PN} &= 3\, \left(\frac{P_\mathrm{b}}{2\pi}\right)^{-5/3}\, \frac{T_\odot^{2/3}M^{2/3}}{1-e^2}  \,,\label{eq:omdot_1PN}\\
\dot{\omega}^\text{2PN} &=  
3\, \left(\frac{P_\mathrm{b}}{2\pi}\right)^{-7/3}\, f_\mathrm{O}\, \frac{T_\odot^{4/3}M^{4/3}}{1-e^2} \,,\label{eq:omdot_2PN}\\
\dot{\omega}^\text{LT}_\mathrm{A} &= -3\, \left(\frac{P_\mathrm{b}}{2 \pi} \right)^{-2} \frac{G}{c^5 T_{\odot} M (1-e^2)^{3/2}} \left(\frac{1}{X_\mathrm{A}} +\frac{1}{3} \right)\, I_\mathrm{A}\, 2\pi\nu_\mathrm{A}\,,
\label{eq:omdot_LT}
\end{align}
where $\nu_\mathrm{A}$ is the spin frequency of pulsar A. 
For the Double Pulsar, the magnitude of the LT contribution is comparable to the 2PN order contribution, but of opposite sign \cite{KW2009, Iorio2009}.

\begin{figure}[t]
    \centering
    \includegraphics[width=0.9\textwidth]{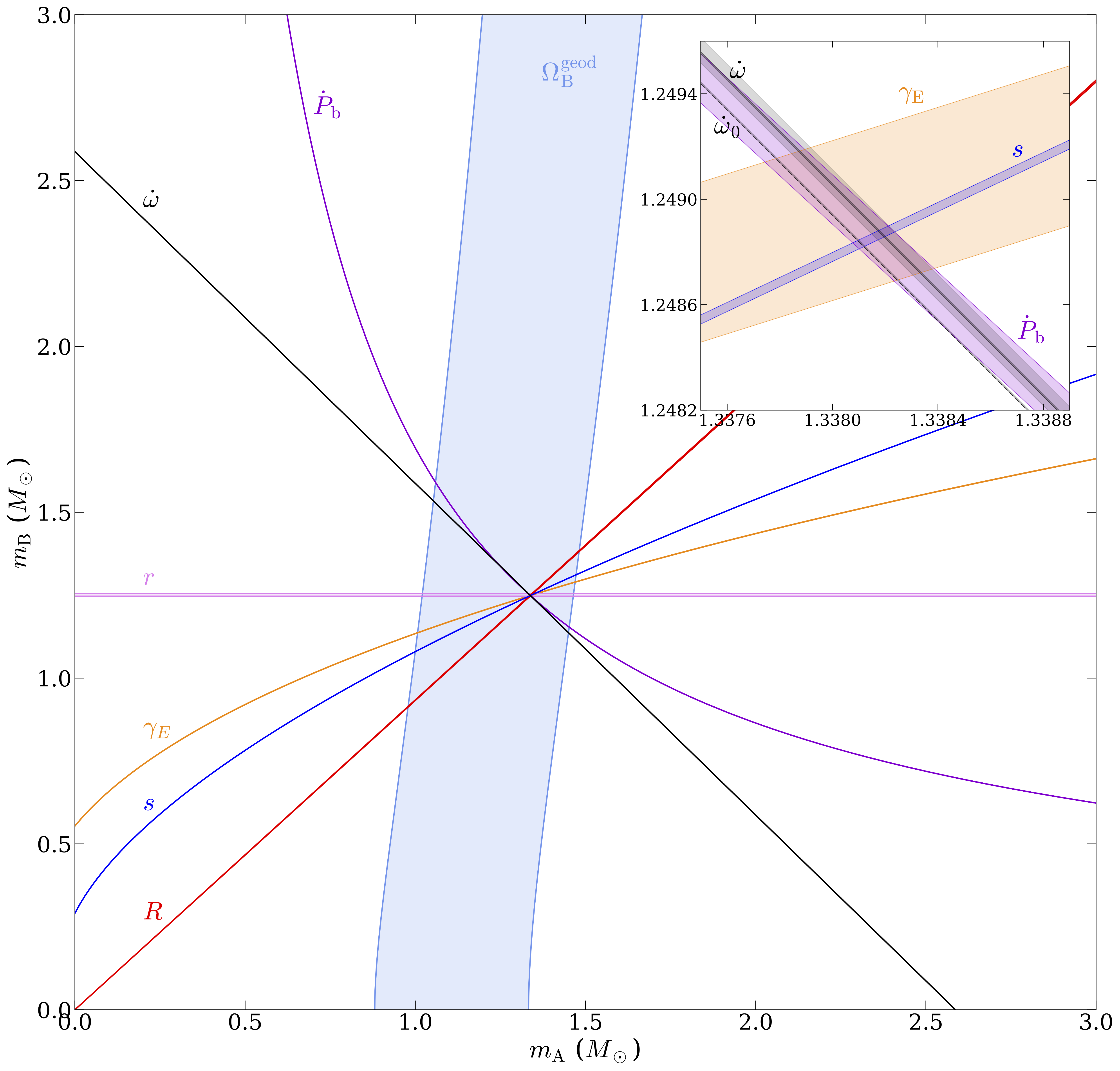}
    \caption{Mass-mass diagram of the Double Pulsar based on the assumption of GR, plotted for six PK parameters and the mass ratio $R$ with pairs of curves showing $\pm 1 \sigma$. The insert on the upper right corner shows an enlarged view of the region of interest, where the influence of the LT effect is shown. The solid black $\dot{\omega}$ line includes LT contribution whereas the dashed $\dot{\omega}_0$ line (i.e. $\dot{\omega}^\text{1PN} +\dot{\omega}^\text{2PN}$) ignores it. Reproduced from Fig.~13 in \cite{Kramer+2021DP} under a Creative Commons Attribution 4.0 International license.}
    \label{fig:mmdiag}
\end{figure}

Fig.~\ref{fig:mmdiag} shows the influence of the LT effect on $\dot{\omega}$. 
If the contribution of $\dot{\omega}^\text{LT}_\mathrm{A}$ can be isolated from the total periastron advance, then $I_\mathrm{A}$ can be measured, as suggested by Refs.~\cite{Lyne2004,Kramer2006}. Even though the uncertainty of the observed advance of periastron is much below the contribution of $\dot{\omega}^\text{LT}_\mathrm{A}$ \cite{Kramer+2021DP}, one has to determine the 1PN and 2PN contributions, which depend on the masses of the two pulsars $m_\mathrm{A}$ and $m_\mathrm{B}$. These masses can be measured very precisely and independently of $\dot{\omega}$ using two other post-Keplerian (PK) parameters: the Shapiro delay parameter $s \equiv \sin{i}$ and the orbital period decay parameter $\dot{P}_\mathrm{b}$ \cite{Hu+2020,Kramer+2021DP}.

Earlier studies on $I_\mathrm{A}$ of PSR~J0737$-$3039A concluded that a $\sim$10\% or higher accuracy MoI measurement is possible by 2030 with the MeerKAT telescope and the Square Kilometre Array (SKA) \cite{KW2009,Kehl}, without accounting for the contribution of spin-down mass-loss of pulsar A to the orbital period derivative. This contribution, however, is proven to be crucial, as it also depends on the MoI \cite{Hu+2020}:
\begin{equation}
    \left(\frac{\dot{P_\mathrm{b}}}{P_\mathrm{b}} \right)^{\dot{m}_\mathrm{A}} = \frac{8\pi^2 G\, I_\mathrm{A} \dot{P}_\mathrm{A} }{c^5 T_{\odot} M {P_\mathrm{A}}^3} \,
    \label{eq:mAdot}
\end{equation}
The quantities $P_\mathrm{A}$ and $\dot{P}_\mathrm{A}$ are the spin period and spin period derivative of pulsar A. 
In addition, the measurement of $\dot{P_\mathrm{b}}$ is contaminated by two kinematic effects that have to be corrected for: the difference in Galactic acceleration of the pulsar system and the Solar System \cite{Nice1995}, and the proper motion of the pulsar, known as Shklovskii effect \cite{Shk70}. Both are proportional to the orbital period $P_\mathrm{b}$.
A good knowledge of the distance and proper motion of the pulsar as well as the Galactic parameters is required (details see e.g. \cite{Hu+2020}). Thus, the contribution that needs to be taken into account in the observed value of $\dot{P_\mathrm{b}}$ is
\begin{align}
    \dot{P}_\mathrm{b}^\text{ obs}
    & = \dot{P}_\mathrm{b}^\text{ GW} + \dot{P}_\mathrm{b}^{\,\dot{m}_\mathrm{A}} + \dot{P}_\mathrm{b}^{\text{ Gal}} + \dot{P}_\mathrm{b}^{\text{ Shk}} \,.
    \label{eq:pbdot}
\end{align}
The leading-order contribution from GW emission enters the equations of motion at 2.5PN order (i.e. $v^5/c^5$)
\citep{Peters1963, Esposito1975, Wagoner1975}:
\begin{align}
\dot{P}_\mathrm{b}^{\text{ 2.5PN}} & = -\frac{192\pi}{5} \, \left(\frac{P_\mathrm{b}}{2\pi}\right)^{-5/3} \frac{T_\odot^{5/3}}{(1-e^2)^{7/2}}\, f_{\rm e} \,\frac{m_\mathrm{A} m_\mathrm{B}}{M^{1/3}} \,,\\
f_{\rm e} & =  1+\frac{73}{24}e^2+ \frac{37}{96}e^4 \,.
\label{eq:pbdot_2p5PN}
\end{align}

Therefore, a self-consistent approach to determining the MoI is to account for the extrinsic contributions and spin-down mass-loss in $\dot{P}_\mathrm{b}$ and solve for the three unknowns ($m_\mathrm{A}, m_\mathrm{B}$, $I_\mathrm{A}$) jointly using  $\dot{P}_\mathrm{b}(m_\mathrm{A}, m_\mathrm{B}, I_\mathrm{A})$, $\dot{\omega}(m_\mathrm{A}, m_\mathrm{B}, I_\mathrm{A})$, and $s\,(m_\mathrm{A}, m_\mathrm{B})$. Based on the timing precision of data from early MeerKAT observations, Ref.~\cite{Hu+2020} simulated timing observations with MeerKAT and the SKA and found that a $\sim$10\% accuracy is likely to be achieved by 2030, given improvements on the Galactic parameters in the coming years. More recent MeerKAT observations, especially those taken in the UHF band, showed an even better timing precision than that used in the simulations \cite{Hu2022}. 
This analysis has been applied to the 16-yr data of PSR~J0737$-$3039A collected from six telescopes without MeerKAT, and an upper limit $I_\mathrm{A} < 3.0 \times 10^{45} \,\mathrm{g \, cm^{2}}$ (90\% confidence) was found \cite{Kramer+2021DP}. The ongoing combination of the 16-yr data set with MeerKAT data is expected to deliver a better constraint on $I_\mathrm{A}$.

As noted in Refs.~\citep{Morrison2004, LS05}, with the mass and spin of the pulsar (which are precisely measured with pulsar timing), even an MoI measurement with $10\%$ accuracy would enable sufficiently accurate estimation of the pulsar radius and provide significant constraints on the EoS of NSs.

%%%%%%%%%%%%%%%%%%%%%%%%%%%%%%
\subsubsection{PSR~J1946+2052, the most compact double neutron star system}
\label{sec:1946}

A DNS system similar to the Double Pulsar is PSR~J1946+2052, which was discovered in PALFA survey with Arecibo \cite{Stovall+2018ApJ}. PSR~J1946+2052 is a 17-ms pulsar in a slightly eccentric ($e=0.06$) 1.88-h orbit, the shortest orbital period among all known DNS systems. 
These features suggest that PSR~J1946+2052 underwent an evolutionary process similar to that of the Double Pulsar, and it resembles the Double Pulsar system after about 40~Myr of further evolution due to GW damping, i.e. this system will merge in approximately 46~Myr.
The faster spin period and smaller orbital period amount to a larger LT contribution $\dot{\omega}^\mathrm{LT}$, making this system a promising candidate for measuring the MoI. 
The recent profile analysis based on FAST data revealed a very small misalignment between the spin vector of this pulsar and the orbital angular momentum vector ($\delta^{\rm SO} \sim 0.2\si{\degree}$) \cite{Meng+2024}. 
Therefore, the LT contribution $\dot{\omega}^\mathrm{LT}$ can be calculated the same way as for the Double Pulsar (see Section~\ref{sec:MoI}) and is shown in Table~\ref{tab:LT}.
The more compact orbit also means a more significant decay of the orbital period due to the emission of GWs; as well as smaller kinematic effects (which scale with $P_\mathrm{b}$), allowing for increased measurement capability. These all suggest that PSR~J1946+2052 is a very promising system for measuring the MoI, especially with the high precision data from FAST.

\begin{figure}[t]
    \centering
    \includegraphics[width=0.85\textwidth]{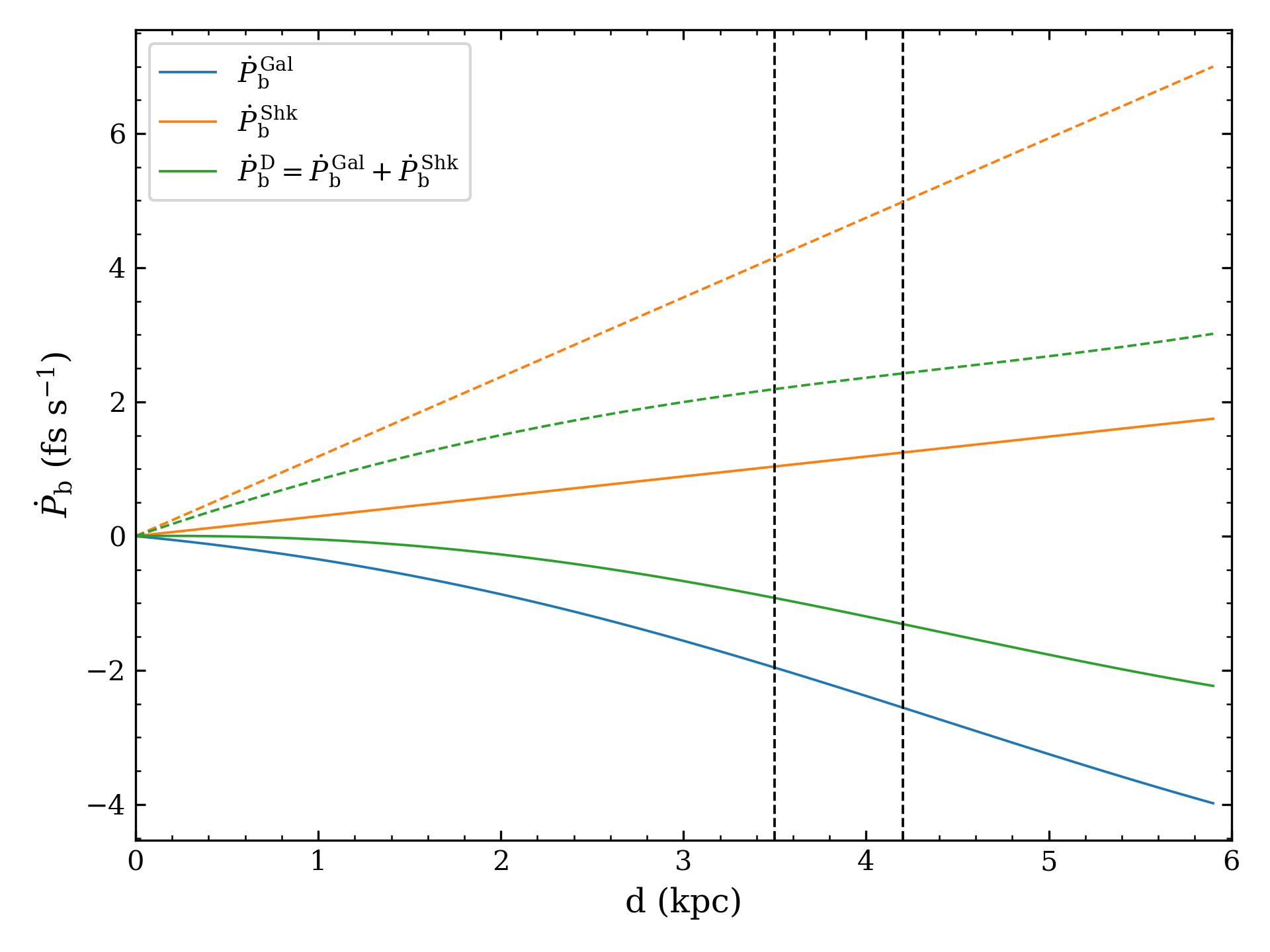}
    \caption{Kinetic contributions of Galactic acceleration (blue) and Shklovskii effect (orange) on the observed orbital period derivative $\dot{P}_\mathrm{b}$ with respect to the distance of PSR~J1946+2052. The total contribution is plotted in green. The solid curves correspond to a moderate proper motion of $\mu_\alpha=\mu_\delta=3\,\mathrm{mas\,yr^{-1}}$, whereas the dashed curves correspond to a proper motion of $\mu_\alpha=\mu_\delta=6\,\mathrm{mas\,yr^{-1}}$. Vertical dashed lines indicate distances of 3.5~kpc and 4.2~kpc derived from the electron-density model NE2001 and YMW16, respectively. }
    \label{fig:xpbdot}
\end{figure}

However, the most challenging part for measuring the MoI of PSR~J1946+2052 is that this system is very far away, making it difficult to correct for the kinematic effects in the $\dot{P_\mathrm{b}}$. Given that the dispersion measure (DM) of PSR~J1946+2052 is $93.965(3) \,\mathrm{pc\,cm^{-3}}$, Ref.~\cite{Stovall+2018ApJ} predicted that the distance of pulsar to be $d=4.2$~kpc and 3.5~kpc based on the electron-density model NE2001 \cite{NE2001} and YMW16 \cite{Yao2017}, respectively. A precise distance measurement using VLBI is unlikely for a faint pulsar at such a distance. Nonetheless, by assuming a proper motion, one could estimate the kinematic contributions over a range of distance as shown in Fig.~\ref{fig:xpbdot}. The green curves indicate the sum of the kinematic effects $\dot{P}_\mathrm{b}^\mathrm{\,D}$, which depend significantly on the proper motion of pulsar (e.g. with $3\,\mathrm{mas\,yr^{-1}}$ and $6\,\mathrm{mas\,yr^{-1}}$), but the difference of $\dot{P}_\mathrm{b}^\mathrm{\,D}$ at $d=3.5$ and $4.2$~kpc changes by only $-0.39$ and $0.23 \,\mathrm{fs\,s^{-1}}$, respectively. With an uncertainty at this level (i.e. only from the uncertainty of distance), the GR's prediction for orbital period decay can still be tested with a better precision than the Hulse–Taylor pulsar \cite{Stovall+2018ApJ}. Therefore, the knowledge of proper motion is essential for the $\dot{P}_\mathrm{b}$ test and MoI measurement, which is expected to come out from future timing.

%%%%%%%%%%%%%%%%%%%%%%%%%%%%%%%%%%%%%%%%
\section{PSR~J0514\texorpdfstring{$-$}{Lg}4002E, a massive binary with a companion in the mass gap}
\label{sec:J0514}

A recent study featured an interesting binary pulsar system discovered by the MeerKAT telescope in the globular cluster NGC~1851, PSR~J0514$-$4002E. The pulsar is in a highly eccentric orbit ($e=0.71$) with a massive compact companion. Based on the measurement of $\dot{\omega}$, the total mass of this system is measured to be $3.8870 \pm 0.0045\, \mathrm{M_\odot}$, which is heavier than any known DNS systems known in the Galaxy \cite{Ferdman+20} and in LIGO events, such as GW190425 \cite{GW190425}. The mass of the companion is between 2.09 and 2.71 $\mathrm{M_\odot}$ (95\% confidence) \cite{Barr2024}, which is larger than any known NS, while being well below the minimum mass observed for BHs in X-ray binaries ($\sim 5 \,\mathrm{M_\odot}$). The nature of the companion can be a very massive NS or a low-mass BH \cite{Barr2024}, or possibly a primordial BH formed during the first-order phase transition in the early Universe \cite{ChenZC+2024}.

If the companion is a massive NS, its mass measurement from future timing can possibly help us to identify the EoS of NS or challenge the current models and provide new insights. 
If the companion is a light BH, its unusually small mass compared to Galactic stellar mass BHs indicates it could be formed from merger of a DNS, such as the case of the GW170817 merger event \cite{LIGO2018}. During such a merger, the BH is likely to acquire a spin angular momentum of $0.6<\chi_\mathrm{c}<0.875$, which would induce a LT precession in $\dot{x} \lesssim 1.7 \times 10^{-13}$ \cite{Barr2024}. In addition, a pulsar with a stellar-mass BH (SBH) companion will allow new tests of gravity theories, such as GR's cosmic censorship hypothesis and no-hair theorem.
It has been shown that with new-generation radio telescopes such as FAST and SKA, a few years of observations on sufficiently compact pulsar-SBH systems will allow precision measurements of the BH mass and spin (1~\% precision) \cite{Liu+2014}, as well as significantly improve the constraint on scalar-tensor theories such as Damour--Esposito-Far\`ese gravity \cite{DEF1993,Batrakov2023}.

%%%%%%%%%%%%%%%%%%%%%%%%%
\section{Future discoveries}
\label{sec:discovery}

The ongoing and future large pulsar surveys with MeerKAT, FAST, and the forthcoming SKA are capable of discovering DNS systems with significantly shorter orbital periods. In fact, a binary pulsar with a 53-min orbital period, PSR~J1953+1844 (M71E), was discovered in 2021 by the FAST Galactic Plane Pulsar Snapshot (GPPS) survey \cite{Pan+2023Natur}. Although with a low-mass companion as a donor star, this system is not of interest for the measurement of LT effect, this discovery certainly adds to our confidence in finding DNS systems with more compact orbits in the near future.

An ideal class of DNS systems should have characteristics similar to PSR~J0737$-$3039A/B and PSR~J1946+2052 (i.e., pulsar spin aligns with the orbital angular momentum), but with much shorter orbital period, in this case the MoI can be determined via the LT precession in the rate of change of periastron $\dot{\omega}$, together with the measurements of the Shapiro parameter $s$ and the orbital period derivative $\dot{P}_\mathrm{b}$. It has been demonstrated in \cite{Hu+2020} that if we were able to discover a 50-min orbit DNS system that resembles the Double Pulsar ($i\sim90\si{\degree}$, similar distance and brightness) and observe it with new-generation telescopes for 10 years, we can expect a measurement of MoI with $\sim 1\%$ precision. This could help enormously in determining the EoS of NSs. In about 10 years, the space GW detectors such as the \emph{Laser Interferometer Space Antenna (LISA)} will be launched and are capable of discovering DNS systems with orbital periods $\sim 10$~min \cite{Lau20}. A multimessenger follow-up strategy with the SKA can deliver better constraints on NS properties \cite{Kyutoku+2019MNRAS,Thrane2020,Hu+2020,Miao+2021ApJ}.

Here, we unveil a streamlined correlation linking the significance of the LT measurement to the orbital period of the DNS system.
As one can see from Eq.~\eqref{eq:pbdot_2p5PN}, the leading contribution from GW emission is related to the orbital period and eccentricity as $\dot{P}_\mathrm{b}^\mathrm{\,GW} \propto {P_\mathrm{b}}^{-5/3} (1-e^2)^{-7/2} f_{\rm e}$, and hence become very large for a short orbital period and a large eccentricity. 
On the other hand, the contributions from Galactic acceleration and Shklovskii effect become smaller as they are proportional to the orbital period, i.e., $\dot{P}_\mathrm{b}^\mathrm{\,Gal+Shk} \propto P_\mathrm{b}$. 
Therefore, the uncertainty of these external effect $\delta\dot{P}_\mathrm{b}^\mathrm{\,Gal+Shk} \propto P_\mathrm{b}$.
We assume that the observed $\dot{P}_\mathrm{b}$ is very accurate so that the dominant uncertainty comes from the observing errors in the external effects (the distance of the pulsar and the Galactic potential measurements), so $\delta\dot{P}_\mathrm{b}^\mathrm{\,GW} \propto \delta\dot{P}_\mathrm{b}^\mathrm{\,Gal+Shk}$, leading to
\begin{equation}
\frac{\delta\dot{P}_\mathrm{b}^\mathrm{\,GW}}{\dot{P}_\mathrm{b}^\mathrm{\,GW}} \propto {P}_\mathrm{b}^{\,8/3} (1-e^2)^{7/2}f_{\rm e}^{-1}\,.
\end{equation}
As the masses are constrained by the intersection of $\dot{P}_\mathrm{b}$ and $s$ lines in a mass-mas diagram like that of Fig.~\ref{fig:mmdiag}, and limited by the less accurate parameter ($\dot{P}_\mathrm{b}$), the uncertainty of the masses $\delta m \propto \delta \dot{P}_\mathrm{b}^\mathrm{\,GW}/\dot{P}_\mathrm{b}^\mathrm{\,GW}$. 
Because $\dot{\omega}$ is nearly parallel to $\dot{P}_\mathrm{b}$ at this parameter space and the LT contribution adds a shift to $\dot{\omega}_0$ towards $\dot{P}_\mathrm{b}$ (see Fig.~\ref{fig:mmdiag}), the uncertainty of the LT contribution $\delta \dot{\omega}^\text{LT}$ is therefore proportional to the uncertainty of masses $\delta m$, and related to the orbital period as $\delta \dot{\omega}^\text{LT} \propto {P_\mathrm{b}}^{8/3} (1-e^2)^{7/2} f_{\rm e}^{-1}$. One can find in Eq.~\eqref{eq:omdot_LT} that the LT contribution is related to the orbital period and eccentricity as $\dot{\omega}^\text{LT} \propto {P_\mathrm{b}}^{-2}(1-e^2)^{-3/2}$, this gives the significance of the LT measurement:
\begin{equation}
    \frac{\dot{\omega}^\text{LT}}{\delta \dot{\omega}^\text{LT}} \propto \frac{{P_\mathrm{b}}^{-2} (1-e^2)^{-3/2} f_{\rm e}}{{P_\mathrm{b}}^{8/3} (1-e^2)^{7/2}} \propto {P_\mathrm{b}}^{-14/3} (1-e^2)^{-5}f_{\rm e}\,.
\end{equation}
Thus for DNS systems with small orbital periods and high eccentricities, measurements of the LT effect can be highly significant and lead to accurate MoI measurements that greatly aid in the constraints on the EoS of dense nuclear matter.

For the case of LT precession in $\dot{x}$, the situation would be a bit more complicated. Unlike a constant LT contribution in $\dot{\omega}$, as shown in Eq.~\eqref{eq:xdot_LT}, $\dot{x}^\text{\,LT}$ follows a cotangent function, exhibiting temporal variability.
However, the small orbital periods help to increase the geodetic spin precession rates enormously, as evidenced by the relationship $\Omega_\text{geod} \propto {P_\mathrm{b}}^{-5/3}$ in Eq.~\eqref{eq:Omega_geo}, resulting in considerably shorter geodetic precession timescales. The measurement of the full cyclical evolution of $x$ and the cyclical evolution of $\omega$, would allow a full determination of the precession angles, thereby allowing a full, precise determination of the spin of the pulsar and its MoI.
Developing a proper orbital model could be useful for this purpose in the near future.

%%%%%%%%%%%%%%%%%%%%%%%%%%
\section{Summary and prospects}
\label{sec:sum}

The extraordinary compactness of NSs makes them fascinating objects to probe the behaviour of matter under extreme conditions that cannot be replicated on Earth. 
This can be achieved by measuring the macroscopic quantities of NSs, namely the mass, radius, tidal deformability, MoI, and quadrupolar moment. 
Since NS masses can be precisely measured via radio pulsar timing, some of the most massive ones have provided crucial constraints on the EOS. However, measuring the NS radius is incredibly challenging. Many of the current methods, such as X-ray observations could suffer from strong model dependencies. In Section~\ref{sec2} we reviewed the method to measure the LT precession, which gives access to the MoI of NSs, and in Section~\ref{sec:evo}, we discussed which types of binary pulsars are suitable for measuring this effect.

In Section~\ref{sec:psrs} we discussed five binary pulsars that have already showed the evidence of LT precession (such as PSR~J1141$-$6545) or are promising to the measurement of this effect, either as a precession of periastron $\dot{\omega}$ or a precession of the projected semi-major axis $\dot{x}$. 
Table~\ref{tab:LT} summarises the estimated values for the LT contributions in four of these systems.
A detailed description of measuring the MoI using $\dot{\omega}$ for DNS systems was given in Section~\ref{sec:MoI}, and the first measurements are expected to come soon from the Double Pulsar, adding complementary constraints to other methods \cite{Hu+2020}.
In Section~\ref{sec:J0514}, we discussed how, depending on the nature of its companion, timing measurements of LT precession in the PSR~J0514$-$4002E system could lead to the measurement of the spin of a BH formed in a merger system.
Such measurements have the potential to shed light on the origins of the BHs in these systems and test the cosmic censorship hypothesis.

As demonstrated in Section~\ref{sec:discovery}, DNS systems with orbital periods shorter than those currently known can yield drastically higher precision MoI measurements, which would greatly enhance our understanding of the behaviour of dense matter.

Such precise measurements of the MoI would represent an exciting development for many reasons besides studies of dense matter. Two are especially relevant here. First, if they are measured for NSs near $1.4\, \rm M_{\odot}$ (such as PSR~J0737$-$3039A and the NSs seen in GW mergers like GW170817), they allow direct comparisons with future improvements of the NS tidal deformability to emerge from ground-based measurements of GW mergers via the I-Love-Q relation. This would represent a fundamental test of our understanding of the structure of compact objects.

Secondly, the MoI of NSs has a very direct application in tests of GR.
As shown in the study with 16-yr Double Pulsar data, the test of the radiative properties in leading order and next-to-leading order of GR also requires the knowledge of the MoI of the pulsar \cite{Hu+2020,Kramer+2021DP}, as the mass loss contribution depends on the MoI and the masses measured from $\dot{\omega}$ and $s$ have to account for the LT precession. 
This calculation currently utilises the multimessenger constraint on the radius of a NS as outlined in \cite{Dietrich+2020Sci} (pink bar in Fig.~\ref{fig:RM}), which are then converted into MoI of pulsar A using the radius-MoI relationship given in \cite{Lattimer2019Univ}. 
Without further improvements in measuring the MoI, the capability to enhance the tests of orbital decay due to GW damping will remain limited. 

Furthermore, the resulting knowledge of the NS structure and the EoS is fundamental for a precise interpretation of tests of alternative theories of gravity done with pulsar timing. In such theories the universality of free fall (UFF) does not apply to massive gravitating objects, because their gravitational properties depend on their internal structure (e.g., \cite{nord_68a,DEF1993}).
This is especially true for very compact objects like NSs; for them the detailed internal structure can only be calculated with the EoS of dense nuclear matter. A violation of the UFF results in dipolar GW emission \cite{Eardley_1975,DEF1993} and a polarisation of the orbit of a binary by a third nearby object, the Nordtvedt effect \cite{nord_68b}; these effects depend not only on the gravity theory but also on the structures of the NSs in the systems.

Several pulsar timing experiments have looked for dipolar GW emission in pulsar-WD systems \cite{Freire_2012,Guo_2021} and the Double Pulsar system \cite{Kramer+2021DP}; other timing experiments attempted to detect the Nordtvedt effect in the motion of the pulsar in the triple system J0337+1715 \cite{Archibald_2018,Voisin_2020}. As predicted by GR, these effects have not been observed, and stringent upper limits on the magnitudes of these effects could be derived. With a precise knowledge of the EoS, we would be able to translate these limits into precise limits on the parameters of alternative gravity theories.

The prospects for the discovery of the types of compact systems mentioned in Section~\ref{sec:discovery} are propitious.
Pulsar surveys with MeerKAT and FAST have already discovered more than 1000 pulsars in the past few years \cite{TRAPUM,FAST_GC,CRAFTS,GPPS}.
Following the successful reception of first light by the SKA prototype dish \cite{SKAMPI} and recent integration of the first SKA dish with MeerKAT \cite{MeerKAT+}, significant progress has been made toward the completion of the SKA, which is expected to be realised in the coming years.
We expect many more exotic compact binary pulsars to be discovered by the SKA, especially DNS systems and pulsar-BH systems with short orbital periods, and monitored with unrivalled precision. Such discoveries, plus the improved timing sensitivity for the currently known binary pulsar systems, will offer exciting opportunity to significantly enhance our ability to measure the MoI of NSs and provide invaluable insights into the fundamental physics governing our universe.

\acknowledgments{We thank the referees for carefully reading the manuscripts and providing valuable comments. We are grateful to Michael Kramer for his helpful comments and Norbert Wex for Figs.~\ref{fig:RM} and \ref{fig:M_MOI}. We also wish to thank Thibault Damour for his important comments. This work is supported by the Max Planck Society.}

\conflictsofinterest{The authors declare no conflict of interest.} 

%%%%%%%%%%%%%%%%%%%%%%%%%%%%%%%%%%%%%%%%%%
%% Optional
%\sampleavailability{Samples of the compounds ... are available from the authors.}

%% Only for journal Encyclopedia
%\entrylink{The Link to this entry published on the encyclopedia platform.}

\abbreviations{Abbreviations}{
The following abbreviations are used in this manuscript:\\

\noindent 
\begin{tabular}{@{}ll}
BH  & Black hole \\
DNS & Double neutron star \\
EoS & Equation of state \\
GR  & General relativity \\
GW  & Gravitational wave \\
LT  & Lense-Thirring \\
MoI & Moment of inertia \\
NS  & Neutron star \\
PSR & Pulsar \\
SBH & Stellar-mass black hole \\
SN  & Supernova \\
UFF & Universality of free fall \\
WD  & White dwarf \\
\end{tabular}
}

%%%%%%%%%%%%%%%%%%%%%%%%%%%%%%%%%%%%%%%%%%
\begin{adjustwidth}{-\extralength}{0cm}

\reftitle{References}


%\bibliography{your_external_BibTeX_file}

%=====================================
% References, variant B: internal bibliography
%=====================================
\bibliography{ref}

% If authors have biography, please use the format below
%\section*{Short Biography of Authors}
%\bio
%{\raisebox{-0.35cm}{\includegraphics[width=3.5cm,height=5.3cm,clip,keepaspectratio]{Definitions/author1.pdf}}}
%{\textbf{Firstname Lastname} Biography of first author}
%
%\bio
%{\raisebox{-0.35cm}{\includegraphics[width=3.5cm,height=5.3cm,clip,keepaspectratio]{Definitions/author2.jpg}}}
%{\textbf{Firstname Lastname} Biography of second author}

% For the MDPI journals use author-date citation, please follow the formatting guidelines on http://www.mdpi.com/authors/references
% To cite two works by the same author: \citeauthor{ref-journal-1a} (\citeyear{ref-journal-1a}, \citeyear{ref-journal-1b}). This produces: Whittaker (1967, 1975)
% To cite two works by the same author with specific pages: \citeauthor{ref-journal-3a} (\citeyear{ref-journal-3a}, p. 328; \citeyear{ref-journal-3b}, p.475). This produces: Wong (1999, p. 328; 2000, p. 475)

%%%%%%%%%%%%%%%%%%%%%%%%%%%%%%%%%%%%%%%%%%
%% for journal Sci
%\reviewreports{\\
%Reviewer 1 comments and authors’ response\\
%Reviewer 2 comments and authors’ response\\
%Reviewer 3 comments and authors’ response
%}
%%%%%%%%%%%%%%%%%%%%%%%%%%%%%%%%%%%%%%%%%%
%\PublishersNote{}
\end{adjustwidth}
\end{document}